\documentclass[lettersize,journal]{IEEEtran}
\usepackage{amsmath,amsfonts}
\usepackage{algorithmic}
\usepackage{algorithm}
\usepackage{array}
\usepackage[caption=false,font=normalsize,labelfont=sf,textfont=sf]{subfig}
\usepackage{textcomp}
\usepackage{amsmath}     
\usepackage{stfloats}
\usepackage{bibentry}
\usepackage{url}
\usepackage{verbatim}
\usepackage{graphicx}
\usepackage{cite}
\usepackage{multirow}
\usepackage{balance}
\usepackage{subfloat}
\usepackage{threeparttable}
\usepackage{makecell}
\usepackage{stfloats}
\usepackage{amssymb}    

\def\BibTeX{{\rm B\kern-.05em{\sc i\kern-.025em b}\kern-.08em
		T\kern-.1667em\lower.7ex\hbox{E}\kern-.125emX}}
\usepackage{balance}
\begin{document}
	\title{Cluster-Based  Time-Variant    Channel \\ Characterization and Modeling for 5G-Railways }
	\author{Xuejian Zhang, ~\IEEEmembership{Student Member,~IEEE,} Ruisi He,~\IEEEmembership{Senior Member,~IEEE,} 
	Bo Ai,~\IEEEmembership{Fellow,~IEEE, }	\\
		Mi Yang,~\IEEEmembership{Member,~IEEE,} 
		Jianwen Ding,~\IEEEmembership{Member,~IEEE,}
		Shuaiqi Gao, Ziyi Qi, Zhengyu Zhang,~\IEEEmembership{Student Member,~IEEE,}
		   and Zhangdui Zhong,~\IEEEmembership{Fellow,~IEEE}
	\thanks{
Part of this paper is presented at the 
International Conference on
Wireless Communications and Signal Processing (WCSP-2024) \cite{zxjwcsp}.

X. Zhang, R. He,  B. Ai, J. Ding, S. Gao, Z. Qi, Z. Zhang, and Z. Zhong are with the School of Electronics and Information Engineering and the Frontiers Science Center for Smart High-speed Railway System, Beijing Jiaotong University, Beijing 100044, China
(email: 23115029@bjtu.edu.cn; ruisi.he@bjtu.edu.cn; boai@bjtu.edu.cn; jwding@bjtu.edu.cn; 23120047@bjtu.edu.cn; 22115006@bjtu.edu.cn; 21111040@bjtu.edu.cn; zhdzhong@bjtu.edu.cn).

M. Yang is with the School of Electronics and Information Engineering and the Frontiers Science Center for Smart High-speed Railway System, Beijing Jiaotong University, Beijing 100044, China, and also with the Henan High-Speed Railway Operation and Maintenance Engineering Research Center, Zhengzhou 451460, China (e-mail: myang@bjtu.edu.cn).


}
}
	\maketitle

\begin{abstract}
With the development of high-speed railways, 5G for Railways (5G-R) is gradually replacing Global System for the Mobile Communications for Railway (GSM-R) worldwide to meet increasing demands. 
The   large bandwidth, array antennas, and  non-stationarity caused by high mobility has made 5G-R channel characterization more complex.
Therefore, it is essential to develop an accurate channel model for 5G-R.
However,  researches on  channel characterization and time-variant models specific to 5G-R frequency bands and scenarios is scarce. 
There are virtually no cluster-based time-variant channel models that capture  statistical properties of 5G-R channel.
In this paper, we propose a   cluster-based  time-variant channel model for 5G-R within an enhanced 3GPP framework, which incorporates time evolution features. 
Extensive channel measurements are conducted on  5G-R private network test line in China. 
We then  extract and analyze typical channel fading characteristics and multipath cluster  characteristics.
Furthermore, birth-death process of the clusters is modeled by using a   four-state Markov chain. 
Finally,  a generalized clustered delay line (CDL) model is established in accordance with 3GPP standard and validated  by comparing the results of measurements and simulations.
This work enhances the understanding of 5G-R channels and presents a flexible cluster-based time-variant channel model. 
The results can be used in the design, deployment, and optimization of 5G-R networks.

\end{abstract}


\section{Introduction}

  \IEEEPARstart{T}{he} 
 railway transportation system is widely acknowledged as an economical, energy-efficient, and effective mode of transporting goods and passengers \cite{he2016}. 
The efficiency and reliability of railway transportation are fundamentally supported by railway wireless communication systems, which are crucial for the reliable transmission of key services and passenger safety \cite{lcz, aibo2020}. 
Over the past few decades, the Global System for Mobile Communications for Railway (GSM-R) has been remarkably successful,  
While  it is a narrowband system with low bandwidth and transmission rates, which no longer suffice to meet the escalating demands for train-ground private network services, such as train multimedia dispatch communications \cite{he2022, he2024, he2023}. 
It is urgently needed to evolve into a new generation  railway communication system. 

With the full commercialization of 5G on public networks, applying 5G technology to railway systems, i.e. 5G for Railways (5G-R), has garnered substantial international interest. 
In Europe, under the auspices of the International Union of Railways (UIC), the Future Railway Mobile Communication System (FRMCS) has introduced 5GRAIL \cite{5GRail} as the successor to GSM-R. 
Railway companies in Japan and South Korea have  deployed 5G infrastructure on select lines and conducted trials to verify communication performance in high-speed scenarios \cite{feng2022}.
China is also shifting its research focus from GSM-R directly to 5G-R without considering Long-Term Evolution for Railway (LTE-R) \cite{chen2018}.
In September 2023, the Ministry of Industry and Information Technology of China authorized the testing of 5G-R private network communication systems, designating the frequency band of 2100 MHz, i.e. uplink  1965-1975 MHz and downlink 2155-2165 MHz \cite{liang2024}. 
It is evident that advancing  railway private network communication system into  5G era is imperative  
and has become a global consensus.

The study of radio wave propagation mechanisms and channel modeling is fundamental to the design and network planning of railway wireless communication systems \cite{zxj2024}. 
Commonly employed link-level channel models include  tapped delay line (TDL) and custered delay line (CDL) models \cite{zzy}. 
For 5G-R system,  communication scenarios  are considerably more complex \cite{he2023}, and the introduction of large bandwidth and multiple-input multiple-output (MIMO)  provides the foundation for developing three-dimension (3D) wideband channel models \cite{ztao2022}. 
More critically,  5G-R channel exhibits pronounced non-stationary characteristics due to  high mobility of trains, which can be characterized by  dynamic birth-death process of multipath clusters \cite{hfy2024}. 
Thus it is imperative to establish a cluster-based  time-variant 5G-R channel model to accurately reflect these characteristics.

Currently,   a  number of wireless channel studies for high-speed railways have emerged.
Typical channel characteristics such as path loss (PL), shadow fading (SF), Rice K-factor based on GSM-R measurements in viaducts and cuttings at 930 MHz, are discussed and modeled in \cite{he2013-2, he2013}. 
Short-term fading behaviors in railway scenarios at 1.89 GHz and 2.605 GHz for LTE-R are analyzed in \cite{zhoutao2019, zt2019-2}. 
A Markov-based multi-link TDL for railway   communications at 460 MHz is established in \cite{zhangbei}.
Refs. \cite{liul2014, liu2014} utilize  measurement data in  viaduct scenarios of  LTE-R system to establish   TDL models based on Markov chain, and \cite{qiu2014} further extract  the inter and intra-cluster parameters in   time domain and delay domain.
Ref. \cite{liang2024} introduces active and passive measurements based on 5G core network at 2100 MHz, discussing channel fading characteristics including PL and root mean square delay spread (RMS DS),
and statistical model of PL for railway 5G marshalling yard scenario is discussed in \cite{djw2023}.
Refs. \cite{hdp} and \cite{lt2022} utilize ray tracing simulators to develop scenarios such as  cutting, viaduct and equipment room at millimeter-wave frequencies and 2100 MHz, respectively, to investigate propagation characteristics.

Despite the considerable research focused on wireless channel  within railway scenarios, most of  existing work focuses on the analysis and modeling of specific channel characteristics, with the majority relying on TDL models.
Although \cite{qiu2014} proposed a CDL model, it is limited to two dimensions and lacks sufficient representation of   angular domain.
What's more, there is basically no research on 5G-R dedicated frequency bands and scenarios.
To the best of our knowledge, the construction of 5G-R dedicated network  is still in its infancy globally, 
dedicated frequency bands have not yet been issued in some countries
and obtaining   5G-R channel measurement data remains highly challenging. 
Consequently, there is currently no established 5G-R time-variant channel model based on measurement data. 

Instead, the more common approach involves employing geometry-based stochastic channel models (GBSMs) to construct CDL \cite{cx2022}.
For instance,  \cite{liu2018} proposed a 3D non-stationary massive MIMO GBSM based on the assumption of uniformly distributed scatterers, and a non-stationary small-scale fading model for 3D MIMO high-speed railway is established in \cite{feng2024}.
Many standardization organizations also have extensively researched on GBSMs to support high-mobility scenarios, such as WINNER II \cite{winner}, 3GPP TR 38.901 \cite{3gpp}, COST 2100 \cite{cost}, QuaDRiGa \cite{qua2014}, and IMT-2020 \cite{imt2020}. 
However, these models still have certain limitations in meeting the specific technical requirements of 5G-R applications. 
For example, WINNER II, 3GPP TR 38.901, and IMT-2020 do not support the  non-stationarity \cite{feng2024}.
While  GMSM-based methods can partially describe  non-stationarity of 5G-R channels, it lacks validation through field measurement data and fails to capture typical channel characteristics. 
Moreover, existing standards are insufficient in fully modeling  channel in typical 5G-R scenarios, particularly with regard to non-stationarity.

In summary, current research on cluster-based 5G-R  time-variant channels is highly inadequate, as reflected in the following aspects. 
First, conducting channel measurements for 5G-R systems is challenging, and the lack of essential measurement data makes it impossible to statistically model typical channel characteristics. 
Second, research on cluster-based 3D statistical channel models for high-speed railways is extremely limited, even beyond 5G-R systems, with minimal focus on 5G-R models that incorporate time-variation. 
Lastly, many standardization organizations have yet to provide detailed classifications for high-speed railway scenarios, particularly 5G-R communication scenarios, and propose time-variant channel models considering non-stationarity for references.

To address the above gaps, we propose a cluster-based time-variant channel model for  5G-R within the 3GPP framework. Extensive channel measurements are conducted, from which we extracted typical 5G-R channel fading and cluster characteristics. 
The accuracy of   proposed 5G-R channel model is subsequently validated against  measured and simulated data.
we have presented some preliminary experimental results in our previous work of \cite{zxjwcsp}.
Specifically, the main contributions of this paper are as follows:
\begin{itemize}
	\item[$\bullet$]  Comprehensive wideband MIMO 5G-R channel measurement campaigns are conducted in  5G-R private network test loop of China, yielding a substantial amount of measurement data. 
	\item[$\bullet$]  A cluster-based time-variant 5G-R channel model is proposed as an enhanced version of  standard 3GPP framework. 
	The model incorporates time evolution and characterizes non-stationarity by modeling  the birth-death process of multipath clusters and dynamically updating cluster parameters.
	\item[$\bullet$]  Typical 5G-R channel fading characteristics and multipath cluster characteristics are extracted and analyzed. Cluster lifetime and a first-order four-state Markov chain are employed to describe the cluster birth-death process.
	\item[$\bullet$]  By comparing  channel characteristic parameters between measurements, simulations and 3GPP model, the model implementation and validation are performed.
\end{itemize}

The remainder of this paper is organized as follows. 
Section II describes the framework of cluster-based 5G-R time-variant model based on 3GPP. 
Measurement system and scenarios are introduced in Section III. 
Then in Section IV, a series of channel characteristics are extracted and analyzed, and cluster characteristics and birth-death process are described in Section V.
Section VI establishes a CDL model based on  3GPP standard and validates the model.
Finally, Section VII draws the conclusions.

\begin{figure*}[!t]
	\centering
	\includegraphics[width=.98\textwidth]{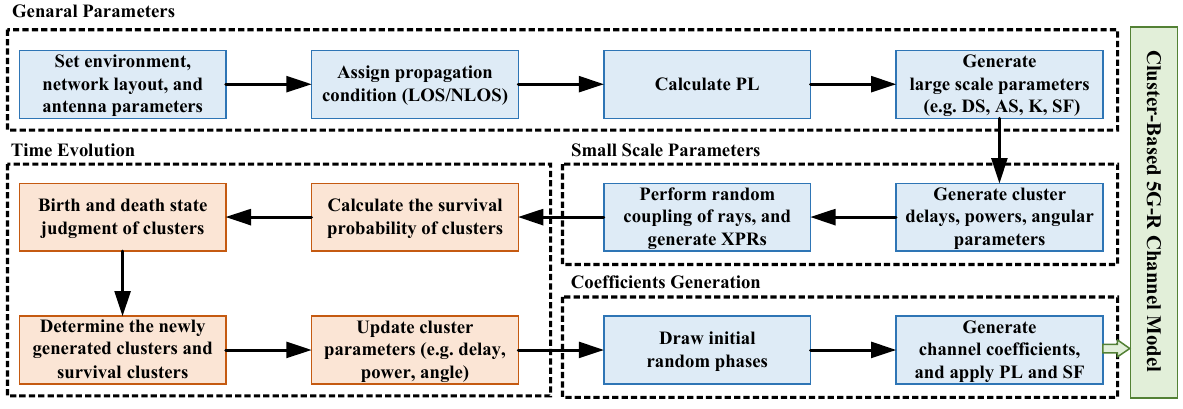}%
	\caption{Framework of non-stationary  5G-R channel modeling.}                    
	\label{framework}
\end{figure*}

\section{Cluster-Based Time-Variant Model for 5G-R Channels}

\subsection{Channel Model  Framework}
Based on  3GPP TR 38.901 \cite{3gpp}, we  propose a generalized 5G-R channel model framework that more comprehensively captures  non-stationary characteristics of wireless channels, 
as illustrated in Fig. \ref{framework}.
The current 3GPP framework does not discuss dynamic channels in detail, especially for smooth evolution of cluster characteristics. 
Although spatial consistency is introduced to make  channel model more reflective of real-world mobile scenarios, it becomes insufficient in realistic dynamic environment with larger spatial and temporal scales \cite{app2019, bian2018, cje}.  
With this in mind, we propose an improved version of  3GPP standard framework, i.e., adding   ``time evolution" part in Fig. \ref{framework}. 
Channel generation procedure in this enhanced framework is  divided into four key parts: general parameters, small-scale parameters, time evolution, and channel coefficient generation.
They are introduced in detail  below.

\textit{1) General Parameters:}
including   initial setup of the whole system, and   generate of corresponding PL model and  large-scale parameters. 
Firstly, types of communication scenario, such as urban micro (UMi) and Rural macro (RMa), network layout and antenna parameters are  chose and set, respectively.
Specially, for base station (BS) and user terminal (UT),  the number of antennas, 3D locations, antenna field patterns, array geometries, azimuth angle ${\phi}$ and elevation angle ${\theta }$  
of each BS and UT in the global coordinate system need  to be considered and determined. 
The speed and direction of UT motion, center frequency  and bandwidth must be set, too. 
Then,  propagation condition, i.e. line of sight (LOS) or non-line of sight  (NLOS)  can be assigned for different BS-UT links.
Finally, the corresponding PL models with the formulas are formed and large scale parameters (LSPs) are calculated, including  DS, AS, K-factor, SF, etc.

\textit{2) Small Scale Parameters:}
including a group of characteristics parameters  of dynamic clusters, such as delays, powers,  angular parameters.
Small scale parameters  of each individual cluster and rays within cluster are generated based on  specific LSPs in general parameters and predefined statistical models in 3GPP \cite{3gpp}. 
Once all the per-ray powers, delays, and angles are obtained, then perform random coupling of rays within a cluster and
apply cross polarization power ratios (XPRs).
Futhermore, the full small scale parameters  of clusters is obtained.
Detailed generation process can be found in \cite{3gpp}.

\textit{3) Time Evolution:}
Non-stationarity of the proposed channel model is embodied in two mechanisms, i.e.,  time-variant parameters and the birth-death process. 
Time-variant parameters are updated constantly and caused by  birth-death process of clusters, while  clusters in a specific scenario can exist over a certain time period, which means the number of  clusters do not change frequently.
Birth-death process  can be modeled in many ways, such as statistical distribution of cluster lifetime and  Markov chains.
Here we follow the suggestions in \cite{chang2023,bian2018,wu2018} to formalize and generalize   birth-death process to obtain  probability of survival or death of each cluster.

Firstly, we define birth-death sampling interval as $ \Delta {t_{BD}}$, and during the 
$ \Delta {t_{BD}}$, cluster birth and death occurs.
Temporal non-stationarity of 5G-R channel is most likely caused by  movement of  UT as mentioned above, and the variable ${\delta _P}\left( {t,\Delta {t_{BD}}} \right)$ is introduced  to describe how much the propagation environment varies during the time interval between $t$  and $t + \Delta {t_{BD}}$.
It represents the sum of  distances traveled by  Tx and Rx from time  $t$ to $t + \Delta {t_{BD}}$ as 
\begin{equation}
	\label{math_K}
	{{\delta _P}\left( {t,\Delta {t_{BD}}} \right) = {v_{UT}}\left( t \right) \cdot \Delta {t_{BD}}},  
\end{equation}
where ${v_{UT}}\left( t \right)$ is the time-variant speed of UT.
Note that, We consider all clusters having the same probability of survival. 
According to  birth-death process, the clusters at time instant $t + \Delta {t_{BD}}$ are considered to be the sum of  clusters that survive from moment $t$ and the clusters that are generated at time intervals ${t_{BD}}$.
The process is determined by a generation rate of clusters ${\lambda _G}$ and a recombination rate of clusters ${\lambda _R}$. 
Consequently, the expectation of total number of clusters in the proposed channel model can be calculated as
\begin{equation}
	\label{math_L}
	{E\left[ {N\left( t \right)} \right] = \frac{{{\lambda _G}}}{{{\lambda _R}}}},  
\end{equation}
where $N\left( t \right)$ represents the time-variant   number of clusters.
The probabilities of clusters at $t + \Delta {t_{BD}}$ survived from $t$ can be modeled as
\begin{equation}
	\label{math_M}
	{{P_{survival}}\left( {\Delta {t_{BD}}} \right) = {e^{ - {\lambda _R} \cdot \frac{{{\delta _P}\left( {\Delta {t_{BD}}} \right)}}{{{D_c}}}}}},  
\end{equation}
where $D_c$ is the scenario dependent correlation factor and typical values   can be chosen in \cite{wu2018}. 
According to Poisson process, the durations between clusters appearance and disappearance follow exponential distribution. 
The expectation of the number of newly generated clusters can be computed as
\begin{equation}
	\label{math_N}
	{E\left[ {{N_{new}}\left( {t + \Delta {t_{BD}}} \right)} \right] = \frac{{{\lambda _G}}}{{{\lambda _R}}}\left( {1 - {e^{ - {\lambda _R} \cdot \frac{{{\delta _P}\left( {\Delta {t_{BD}}} \right)}}{{{D_c}}}}}} \right)}.  
\end{equation}

In our proposed channel model, the disappearing clusters at each time instant are removed, while for the newly generated clusters, delays,  powers  and angle parameters are randomly generated which are similar to the previous steps. 
For the surviving clusters from   previous time instant, the update process of delays, powers and angle parameters will be described in rest of this section.

\textit{4) Coefficients Generation:}
Draw random initial phase  $\left\{ {\Phi _{n,m}^{\theta \theta },\Phi _{n,m}^{\theta \phi },\Phi _{n,m}^{\phi \theta },\Phi _{n,m}^{\phi \phi }} \right\}$ for each ray $m$ of each cluster $n$ and for four different polarisation combinations $\left\{ {\theta \theta ,\theta \phi ,\phi \theta ,\phi \phi } \right\}$, which are uniformly distributed in the range $\left( { - \pi ,\pi } \right)$.
Then channel impulse responses (CIRs)  between $u$th UT and $s$th BS are generated  based on 3GPP TR  38.901,  which consists of two components, i.e.,  LOS component and  NLOS component.
For NLOS condition,  CIR is
\begin{equation}
	\label{math_O}
	{\begin{array}{l}
			H_{u,s,n,m}^{NLOS}\left( t \right) = \sqrt {\frac{{{P_n}}}{M}} {\left[ {\begin{array}{*{20}{c}}
						{{F_{rx,u,\theta }}\left( {{\theta _{n,m,EOA}},{\phi _{n,m,AOA}}} \right)}\\
						{{F_{rx,u,\phi }}\left( {{\theta _{n,m,EOA}},{\phi _{n,m,AOA}}} \right)}
				\end{array}} \right]^T}\\
			\left[ {\begin{array}{*{20}{c}}
					{\exp \left( {j\Phi _{n,m}^{\theta \theta }} \right)}&{\sqrt {{\kappa _{n,m}}^{ - 1}} \exp \left( {j\Phi _{n,m}^{\theta \phi }} \right)}\\
					{\sqrt {{\kappa _{n,m}}^{ - 1}} \exp \left( {j\Phi _{n,m}^{\phi \theta }} \right)}&{\exp \left( {j\Phi _{n,m}^{\phi \phi }} \right)}
			\end{array}} \right]\\
			\left[ {\begin{array}{*{20}{c}}
					{{F_{tx,s,\theta }}\left( {{\theta _{n,m,EOD}},{\phi _{n,m,AOD}}} \right)}\\
					{{F_{tx,s,\phi }}\left( {{\theta _{n,m,EOD}},{\phi _{n,m,AOD}}} \right)}
			\end{array}} \right]\exp \left( {j2\pi \frac{{\hat r_{rx,n,m}^T \cdot {{\bar d}_{rx,u}}}}{{{\lambda _0}}}} \right)\\
			\exp \left( {j2\pi \frac{{\hat r_{tx,n,m}^T \cdot {{\bar d}_{tx,s}}}}{{{\lambda _0}}}} \right) \cdot \exp \left( {j2\pi \frac{{\hat r_{rx,n,m}^T \cdot {{\bar v}_{UT}}}}{{{\lambda _0}}}t} \right)
		\end{array}},  
\end{equation}
where $F_{rx,u,\theta }$, $F_{rx,u,\phi }$, $F_{tx,s,\theta }$, $F_{tx,s,\phi }$ are the field patterns of the $u$th UT and $s$th BS in the direction of  spherical basis vector,
respectively;
$\kappa _{n,m}$ is the XPR for  $m$th ray of $n$th cluster;
$\hat r_{rx,n,m} $ and  $\hat r_{tx,n,m} $ denote the spherical unit vector with its corresponding azimuth arrival angle and elevation arrival angle, 
while ${\bar d}_{tx,s} $ and ${\bar d}_{rx,u}$ denote the location vector at the antennas of BS and UT  respectively;
$\lambda _0$ is the wavelength of carrier frequency and $M$ is the number of rays within the cluster;
${\bar v}_{UT}$ is  the timeinvariant velocity of UT.

For LOS condition, CIR of the LOS path is given as
\begin{equation}
	\label{math_P}
	{\begin{array}{l}
			H_{u,s,1}^{LOS}\left( t \right) = {\left[ {\begin{array}{*{20}{c}}
						{{F_{rx,u,\theta }}\left( {{\theta _{LOS,EOA}},{\phi _{LOS,AOA}}} \right)}\\
						{{F_{rx,u,\phi }}\left( {{\theta _{LOS,EOA}},{\phi _{LOS,AOA}}} \right)}
				\end{array}} \right]^T}\\
			\left[ {\begin{array}{*{20}{c}}
					1&0\\
					0&{ - 1}
			\end{array}} \right]\left[ {\begin{array}{*{20}{c}}
					{{F_{tx,s,\theta }}\left( {{\theta _{LOS,EOD}},{\phi _{LOS,AOD}}} \right)}\\
					{{F_{tx,s,\phi }}\left( {{\theta _{LOS,EOD}},{\phi _{LOS,AOD}}} \right)}
			\end{array}} \right]\\
			\exp \left( {j2\pi \frac{{\hat r_{rx,LOS}^T \cdot {{\bar d}_{rx,u}}}}{{{\lambda _0}}}} \right)\exp \left( {j2\pi \frac{{\hat r_{tx,LOS}^T \cdot {{\bar d}_{tx,s}}}}{{{\lambda _0}}}} \right)\\
			\exp \left( {j2\pi \frac{{\hat r_{rx,LOS}^T \cdot {{\bar v}_{UT}}}}{{{\lambda _0}}}t} \right)\exp \left( { - j2\pi \frac{{{d_{3D}}}}{{{\lambda _0}}}} \right)
		\end{array}},  
\end{equation}
where $d_{3D}$ is the 3D distance between BS and UT.
Note that, CIR of LOS condition is the sum of CIRs of LOS path and NLOS condition with the scaling based on K-factor $K_{R}$, given by
\begin{equation}
	\label{math_Q}
	{\begin{array}{l}
			H_{u,s}^{LOS}\left( {\tau ,t} \right) = \sqrt {\frac{1}{{{K_R} + 1}}} H_{u,s,n,m}^{NLOS}\left( {\tau ,t} \right)\\
			\qquad \qquad \quad + \sqrt {\frac{{{K_R}}}{{{K_R} + 1}}} H_{u,s,1}^{LOS}\left( t \right)\delta \left( {\tau  - {\tau _1}} \right)
		\end{array}},  
\end{equation}
where $\delta \left(  \cdot  \right)$  is the Dirac's delta function.
The LOS component is associated with the shortest possible delay of all $\tau _n$, i.e., $\tau _1$.
As final step,  path loss and shadowing are  applied for all  channel coefficients.


\subsection{Cluster Parameters Evolution}
On a time scale, cluster parameters will be dynamically updated due to the movement of BS or UT. 
To fully characterize   non-stationarity of  dynamic channel, it is particularly important to be able to update  time-variant cluster parameters \cite{hrs2021}.
Note that, this paper does not study the scenario of train-to-train communications.  
Therefore, for typical high-speed railway scenarios,  BS is fixed, and scatterers around the railway, such as high-rise buildings, mountains,  cuts, can also be considered fixed. 
Pedestrians and regular vehicles rarely appear around railways, and their impact on 5G-R channels is relatively small.
Given the specific characteristics of 5G-R railway communication, and drawing on the standards established by 3GPP \cite{3gpp}, IMT-2020 \cite{imt2020},  WINNER \uppercase\expandafter{\romannumeral2} \cite{winner}, as well as previous work on non-stationary channel models \cite{chang2023,bian2018,app2019}, we model the update process of cluster delay, power, and angle parameters.

\textit{1) Update Delays, ${\tau _n}\left( t \right)$.}
At moment ${{t_k} = {t_{k - 1}} + \Delta t}$, the delay of the $n$th cluster is given as
\begin{equation}
	\label{math_a}
	{\widetilde {{\tau _n}}\left( {{t_k}} \right) = \widetilde {{\tau _n}}\left( {{t_{k - 1}}} \right) - \frac{{{{\widehat r}_{rx,n}}{{\left( {{t_{k - 1}}} \right)}^T}\overline v \left( {{t_{k - 1}}} \right)}}{c}\Delta t},  
\end{equation}
where $c$ is the speed of light, $\overline v \left( {{t_{k - 1}}} \right)$ is  UT velocity vector in 3D and given as
\begin{equation}
	\label{math_b}
	{\overline v \left( {{t_{k - 1}}} \right) = {\left[ {\begin{array}{*{20}{c}}
					{{V_X}\left( {{t_{k - 1}}} \right)}&{{V_Y}\left( {{t_{k - 1}}} \right)}&{{V_Z}\left( {{t_{k - 1}}} \right)}
			\end{array}} \right]^T}}.  
\end{equation}
After updating the delays according to Equation \eqref{math_a}, the delays over the mobility range are normalized. 
Thus the final delay set is
\begin{equation}
	\label{math_c}
	{{\tau _n}\left( {{t_k}} \right) = {\tilde \tau _n}\left( {{t_k}} \right) - \min \left\{ {{{\tilde \tau }_n}\left( {{t_k}} \right)} \right\}},  
\end{equation}
where $t_{k}$ covers the entire duration of dynamic channel.

\textit{2) Update Powers, ${P _n}\left( t \right)$.}
After obtaining   delay of   update completion of  the $n$th cluster at moment $t_k$,   power update only needs to substitute Equation \eqref{math_c} into 
\begin{equation}
	\label{math_C}
	{P_n^ *  = \exp \left( { - {\tau _n}\frac{{{r_\tau } - 1}}{{{r_\tau } \cdot DS}}} \right) \cdot {10^{ - {Z_n}/10}}},  
\end{equation}
where ${P _n}$  denotes cluster power, ${Z_n} \sim N\left( {0,{\zeta ^2}} \right)$, $\zeta$ represents the per-cluster shadowing and can be found in \cite{3gpp}.

\textit{3) Update Angle Parameters.}
For the $n$th cluster, define rotation matrix $R$ to transfer $ {{\overline v \left( {{t_{k - 1}}} \right)}} $ to ${\bar v^ * }\left( {{t_k} - 1} \right)$:
\begin{equation} \label{math_d}
	\begin{aligned}
		{{\bar v}^ * }\left( {{t_k} - 1} \right) &= R \cdot \bar v\left( {{t_k} - 1} \right)\\
		&= {\left[ {\begin{array}{*{20}{c}}
					{V_X^ * \left( {{t_{k - 1}}} \right)}&{V_Y^ * \left( {{t_{k - 1}}} \right)}&{V_Z^ * \left( {{t_{k - 1}}} \right)}
			\end{array}} \right]^T},\\
	\end{aligned}
\end{equation}
where the definitions of $R$  in the standards are different for LOS   and NLOS clusters, and a detailed introduction can be found in \cite{3gpp}.
Now, Cluster departure angles and arrival angles in radians are updated as
\begin{equation}
	\label{math_d}
	{\begin{array}{l}
			{\phi _{n,AOD}}\left( {{t_k}} \right) = {\phi _{n,AOD}}\left( {{t_{k - 1}}} \right) + \\
			\dfrac{{{{\bar v}^ * }{{\left( {{t_{k - 1}}} \right)}^T} \cdot \hat \phi \left( {{\theta _{n,EOD}}\left( {{t_{k - 1}}} \right),{\phi _{n,AOD}}\left( {{t_{k - 1}}} \right)} \right)}}{{c \cdot {{\tilde \tau }_n}\left( {{t_{k - 1}}} \right) \cdot \sin \left( {{\theta _{n,EOD}}\left( {{t_{k - 1}}} \right)} \right)}}\Delta t
	\end{array}},  
\end{equation}

\begin{equation}
	\label{math_e}
	{\begin{array}{l}
			{\theta _{n,EOD}}\left( {{t_k}} \right) = {\theta _{n,EOD}}\left( {{t_{k - 1}}} \right) + \\
			\dfrac{{{{\bar v}^ * }{{\left( {{t_{k - 1}}} \right)}^T} \cdot \hat \theta \left( {{\theta _{n,EOD}}\left( {{t_{k - 1}}} \right),{\phi _{n,AOD}}\left( {{t_{k - 1}}} \right)} \right)}}{{c \cdot {{\tilde \tau }_n}\left( {{t_{k - 1}}} \right)}}\Delta t
	\end{array}}, 
\end{equation}

\begin{equation}
	\label{math_f}
	{\begin{array}{l}
			{\phi _{n,AOA}}\left( {{t_k}} \right) = {\phi _{n,AOA}}\left( {{t_{k - 1}}} \right) + \\
			\dfrac{{{{\bar v}^ * }{{\left( {{t_{k - 1}}} \right)}^T} \cdot \hat \phi \left( {{\theta _{n,EOA}}\left( {{t_{k - 1}}} \right),{\phi _{n,AOA}}\left( {{t_{k - 1}}} \right)} \right)}}{{c \cdot {{\tilde \tau }_n}\left( {{t_{k - 1}}} \right) \cdot \sin \left( {{\phi _{n,EOA}}\left( {{t_{k - 1}}} \right)} \right)}}\Delta t
	\end{array}}, 
\end{equation}

\begin{equation}
	\label{math_g}
	{\begin{array}{l}
			{\theta _{n,EOA}}\left( {{t_k}} \right) = {\theta _{n,EOA}}\left( {{t_{k - 1}}} \right) + \\
			\dfrac{{{{\bar v}^ * }{{\left( {{t_{k - 1}}} \right)}^T} \cdot \hat \theta \left( {{\theta _{n,EOA}}\left( {{t_{k - 1}}} \right),{\phi _{n,AOA}}\left( {{t_{k - 1}}} \right)} \right)}}{{c \cdot {{\tilde \tau }_n}\left( {{t_{k - 1}}} \right)}}\Delta t
	\end{array}}, 
\end{equation}
where ${\hat \theta \left( {\theta ,\phi } \right)}$ and $\hat \phi \left( {\theta ,\phi } \right)$ are the spherical unit vectors defined in \cite{3gpp} as
\begin{equation}
	\label{math_h}
	{\hat \theta  = \left[ {\begin{array}{*{20}{c}}
				{\cos \theta \cos \phi }\\
				{\cos \theta \sin \phi }\\
				{ - \sin \theta }
		\end{array}} \right],\hat \phi  = \left[ {\begin{array}{*{20}{c}}
				{ - \sin \phi }\\
				{ + \cos \phi }\\
				0
		\end{array}} \right].} 
\end{equation}

So far, through the above steps we  complete the update of  surviving cluster parameters   within  time interval $ \Delta {t_{BD}}$. 
For the newly generated clusters, their parameters are randomly generated according to the generation process of small scale parameters in Section II-(A).

 \begin{figure}[!t]
	\centering
	\includegraphics[width=.48\textwidth]{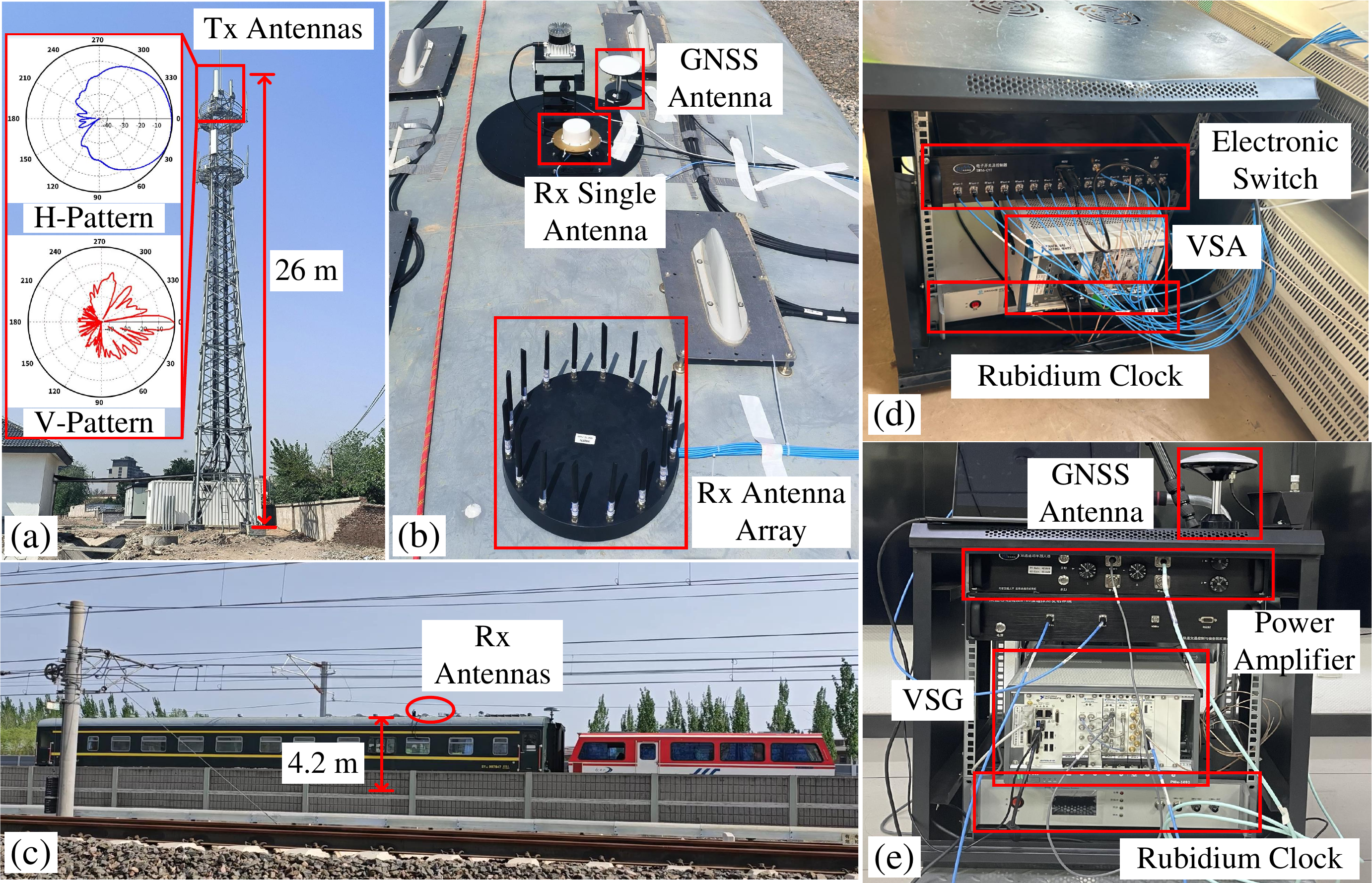}%
	\caption{Measurement Equipments.
		(a)   BS and Tx antennas;
		(b) On–board roof Rx antennas;
		(c) Test train;
		(d) Primary equipments of Tx;
		(e) Primary equipments of Rx.
	}                    
	\label{device}
\end{figure}

 \begin{figure}[!t]
	\centering
	\includegraphics[width=.48\textwidth]{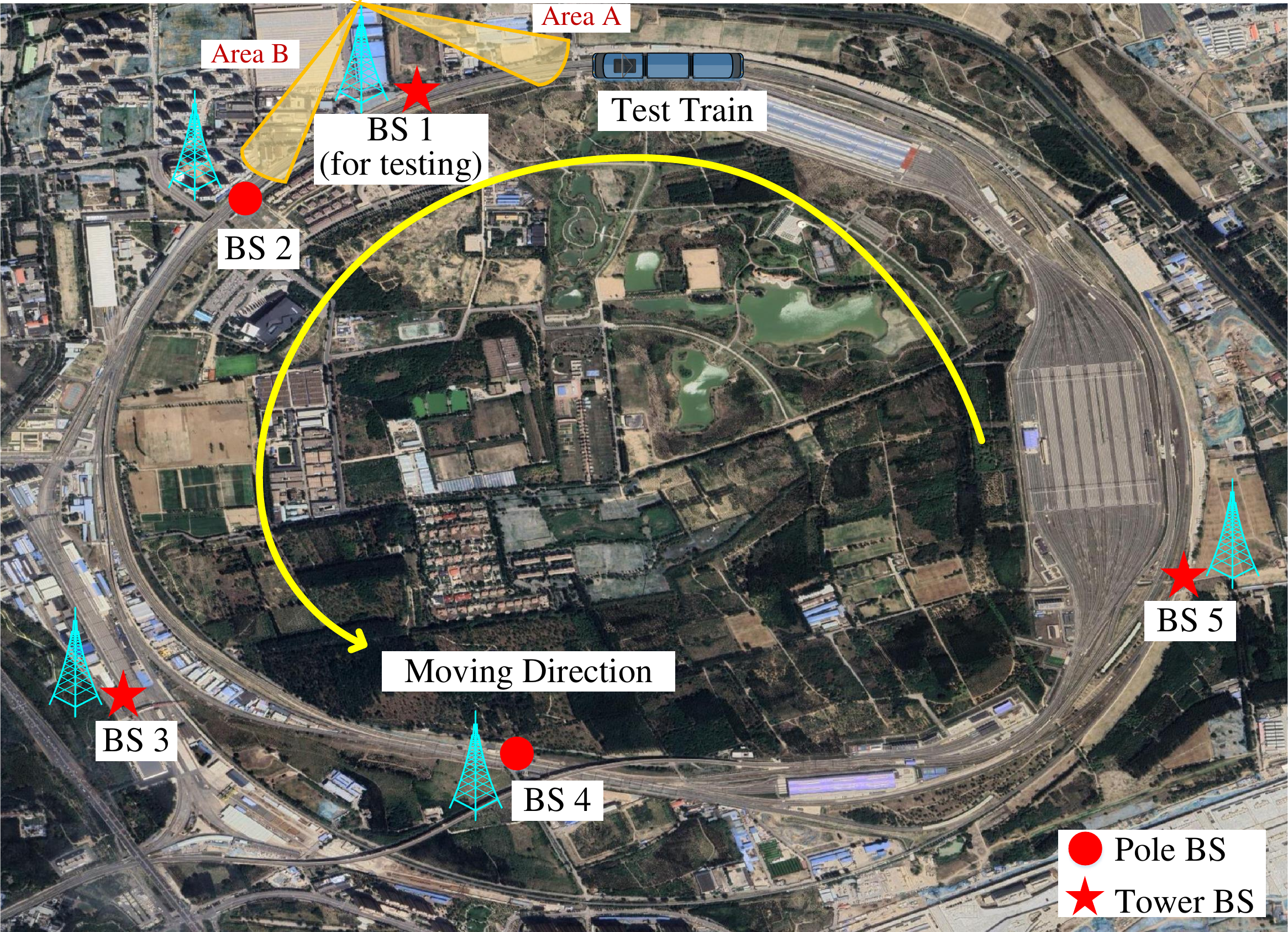}%
	\caption{Measurement scenario: loop railway test line.}                    
	\label{Scenarios}
\end{figure}

\section{Measurement Campaign}  

\subsection{Measurement System}
The key equipment  in  the measurements is shown in Fig. \ref{device}. 
The measurement system comprises transmitting (Tx) and receiving (Rx) subsystems, with the core components being a vector signal generator (VSG) and a vector signal analyzer (VSA). 
Both VSG and VSA are based on National Instruments (NI) equipment.
Specifically, NI PXIe-5673, NI PXIe-5663, and USRP 2954 are utilized as VSG and VSAs,  respectively.
Tx antennas are $\pm\ 45^{\circ}$ polarized directional panel antenna with  gain of 17.5 dBi. 
They are installed on a tower with  height of 26 m, as shown in Fig. 3 (a), which also contains antenna patterns. 
To realize  measurements of small-scale fading and angle domain at the same time,
16-element circular array and single conical antenna are deployed   at Rx and controlled by NI PXIe-5663 and USRP-2954, respectively, both of which are vertically polarized omnidirectional antennas with  gain of 3 dBi. 
They can simultaneously receive multi-antenna data with 16 snapshots/s and single-antenna data with   500 snapshots/s.
Rx antennas are magnetically attached to the top of  test train at  height of   4.2 m, as depicted  in Figs. \ref{device}(b) and (c).

\begin{table}[!t]
	\renewcommand{\arraystretch}{1.5}
	\begin{center}
		\caption{Configuration of 5G-R Channel Measurement System.} 
		\label{measure_system}
		\begin{tabular}{c| c}
			\hline
			\hline
			Parameters            & Description \\  
			\hline
			Center Frequency      & 2160 MHz \\
			\hline
			Bandwidth             &     10 MHz     \\
			\hline
			Number of frequency points & 513  \\
			\hline
			Tx antennas           &  $\pm\ 45^{\circ}$ directional panel antennas \\
			\hline
			Rx antennas           &  \makecell[c]{Vertical polarized omni-directional \\ 16-element circular array \\  and  biconical antenna} \\
			\hline
			Transmitting Power    &  43 dBm   \\
			\hline
			Sampling rate         & \makecell[c]{500 snapshots/s with single antenna \\ (one snapshot per 0.32$\lambda$) \\ 16 snapshots/s with  circular array \\ (one snapshot per 10$\lambda$) }   \\
			
			\hline
			Height of  antenna   & \makecell[c]{26 m (Tx antennas)  \\ 4.2 m (Rx antennas)} \\
			\hline
			Antenna gain   & \makecell[c]{17.5 dBi (Tx antennas)  \\ 3 dBi (Rx antennas)} \\
			\hline
			Length of   loop line         & 9 km   \\
			\hline
			BS feeder loss     &   4.1 dB  \\
			\hline
			Speed of locomotive   & 80 km/h  \\
			\hline
			\hline
		\end{tabular}
	\end{center}
\end{table}

Except antennas, other key equipment is housed in two 12U cabinets, as shown in Figs. \ref{device}(d) and (e). 
Rx cabinet is placed inside   test train, 
while Tx cabinet is located in  equipment room under the tower.
Power amplifier provides a maximum gain of 43 dB with an adjustable step of 1 dB. 
Electronic switch enables millisecond-level fast switching between 16 sub-channels  
to control  circular array, implementing single-transmit multiple-receive (SIMO) based on time division multiplexing.
To ensure time synchronization, two rubidium clocks tamed by Global Navigation Satellite System (GNSS) signals are  employed.
Additionally,  rubidium clocks can also output longitude and latitude coordinates, allowing us to obtain real-time position information of Tx and Rx.
Detailed measurement system configuration is shown in Table \ref{measure_system}.

%

%

\subsection{Measurement Scenario}
Measurement campaign is carried out at the National Railway Track Test Center, located  northeast of Beijing, China.
It features the largest loop railway test line in Asia, spanning a total length of approximately 9 km \cite{liang2024}.
Along this railway,  five 5G-R dedicated network base stations (BSs) have been constructed, equipped with BS equipment from multiple manufacturers, which are illustrated in  Fig. \ref{Scenarios}. 

During the measurements,  Tx cabinet is connected to  Tx antennas of BS 1 to radiate vector signals outward, with  carrier frequency of 2160 MHz and  bandwidth of 10 MHz.
This setup complies with  5G-R dedicated test frequency band issued by the Ministry of Industry and Information Technology of China. 
It is important to note that only two Tx antennas with opposite radiation directions are used to cover Areas A  and B, as shown in Fig. \ref{Scenarios}.
It can be seen that in Area A, there are basically no abundant residential buildings, only a few large facilities along the railway, such as  train control center, operation and maintenance building,  and the scatterers are relatively sparse; 
while in Area B,  more residential houses and low buildings are present,  the scatterers are denser.
What's more, to eliminate signal interference from public network and inter-station,  frequency band of 2155-2165 MHz is cleared, and all   BSs except BS 1 are   shut down. 

Test train consists of a locomotive and a carriage, as shown in Fig. \ref{device}(c).
It maintains a speed of 80 km/h and moves counterclockwise along the circular railway line for several laps to ensure that sufficient channel data is obtained.
The measured environment can be classified as a rural area, characterized by surrounding trees and sparse low-rise buildings. 
Additionally, specific railway objects, such as low partition walls and contact network poles along the railway, are present in the measurement environment.

\subsection{System Calibration and Data Processing}
Measurement system calibration is a crucial step in ensuring accurate results.
It  consists of back-to-back measurement and antenna calibration.
Back-to-back measurement refers to directly connecting   Tx and Rx via a radio frequency  cable and several attenuators without involving  wireless channel.
It effectively compensates for   amplitude-frequency response  caused by cables, adapters, transceivers, etc.
Moreover, Tx and  Rx antennas have been accurately measured in an anechoic chamber for radiation patterns.
This allows for the correction of errors introduced by antenna radiation characteristics during data processing, known as antenna calibration.

Channel impulse response $ h(t,\tau ) $ can be obtained by calculating   channel transfer function in   frequency domain after considering system calibration and then performing a Fourier transform on it, where $t$ is  index of measurement time snapshot, $ \tau $ is   delay bin. 
Detailed description can be found in \cite{yangmi2019}.


\begin{figure}[!t]
	\centering
	
	\subfloat[]{\includegraphics[width=.46\textwidth]{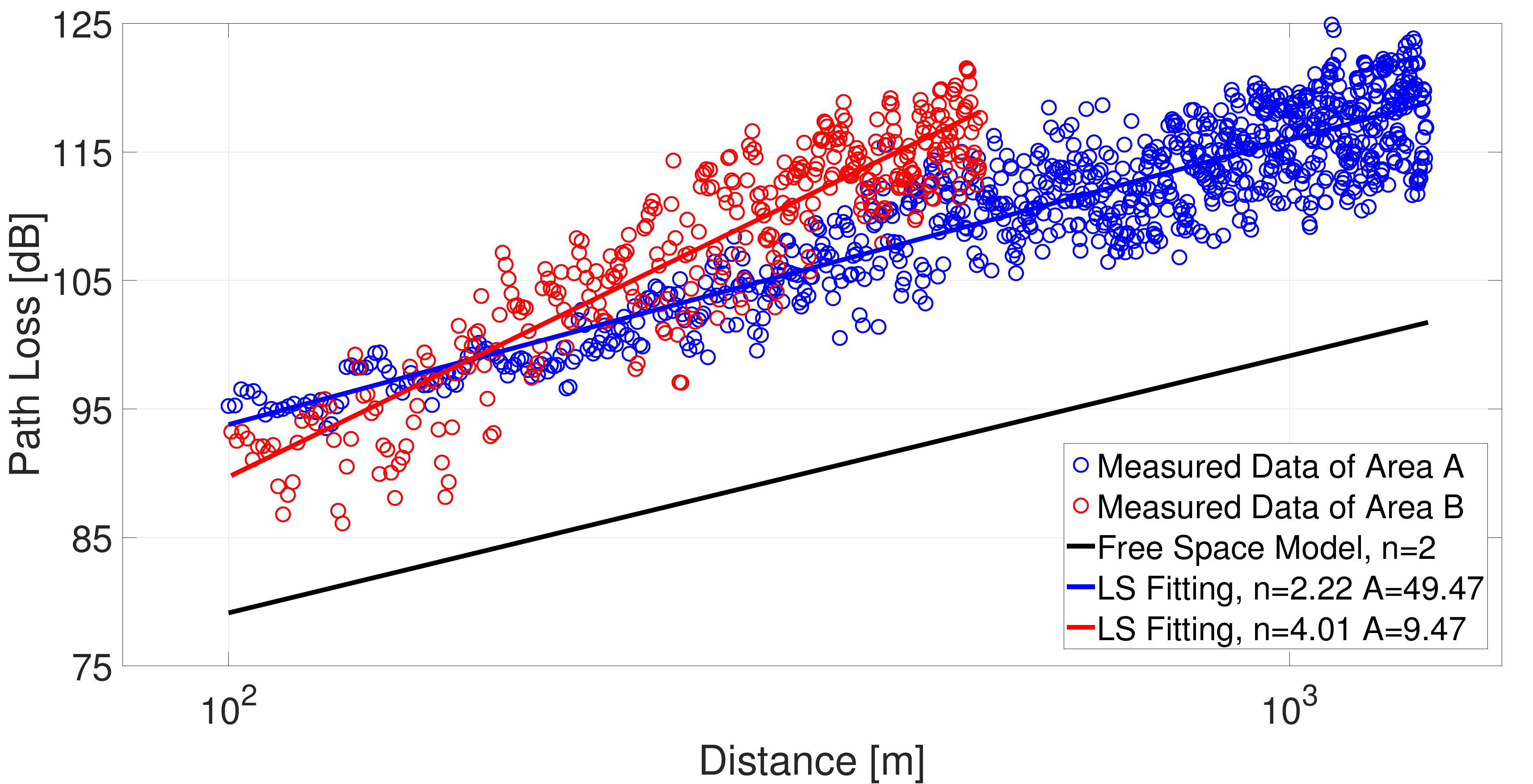}%
		\label{PL}}
	\quad
	\subfloat[]{\includegraphics[width=.46\textwidth]{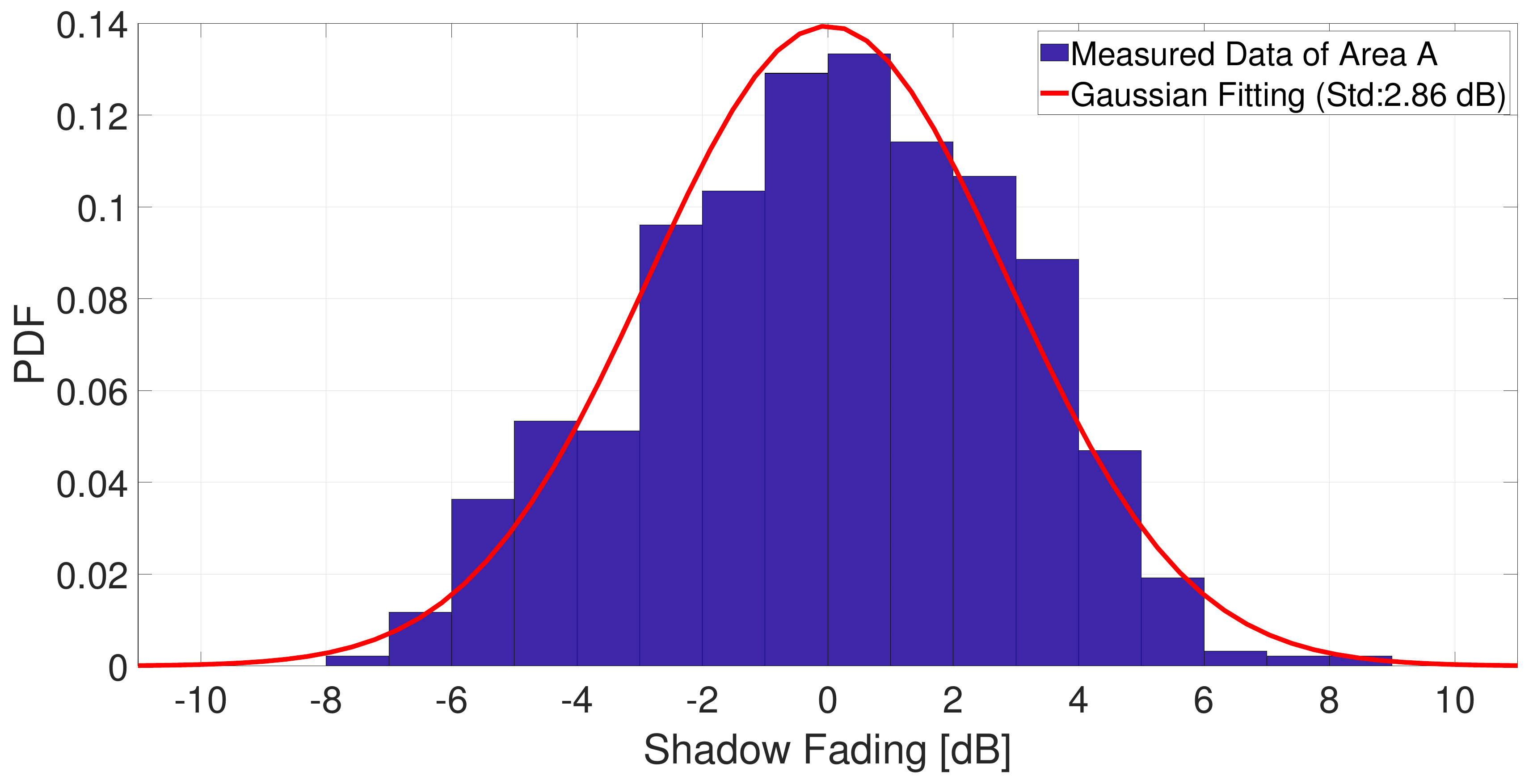}%
		\label{SF}}
	\caption{
		(a) Measured PL with the corresponding Log-distance fitting.
		(b) Measured SF and Gaussian fitting.
	}
	\label{PL-SF}
\end{figure}

\section{Channel  Characteristics}
In this section, we extract and analyze typical channel characteristics based on single-antenna measured data, such as PL, SF, power delay profile (PDP), RMS DS,  K-factor, and stationary region.

\subsection{Path Loss and Shadow Fading}
PL and SF are large-scale propagation characteristics, 
which can be determined by averaged channel gain as follows

\begin{equation}
	\label{math_4}
	L =  - 10{\log _{10}}(\frac{1}{W}\sum\limits_{T = t - \frac{W}{2}}^{t + \frac{W}{2} - 1} {\sum\limits_{\tau  = 1}^{{N_f} } {{{\left| {h(T,\tau )} \right|}^2}} } ).
\end{equation}
where ${N_f}$ is the number of measured frequency points, $L$ is large-scale components in dB-scale including PL and SF,
and $W$ is a $40\lambda$ sliding window, $\lambda$ denotes wavelength.
After eliminating small-scale fading by a sliding window of $40\lambda$, the dB-valued PL and SF can be generally modeled as \cite{zxj2023}
\begin{equation}
	\label{math_5}
	L(d) = A + 10n{\log _{10}}(\frac{d}{{{d_0}}}) + {X_\sigma },
\end{equation}
where $A$ is intercept value and $n$ is path loss exponent, $d$ and $d_0$ represents the  distance between Tx and Rx and the reference distance which set to 1 m, respectively. 
SF is represented by $ X_\sigma $, which is a zero-mean Gaussian distributed random variable with standard deviation $ \sigma $.

The measured distance-dependent path loss and the corresponding linear fitting with Least Square (LS) regression in Areas A and B  are shown in Fig. \ref{PL-SF}(a).
For eliminating near-field effect caused by antenna directivity, we ignore the data within first 100 m.
The observed disparity in path loss between Areas A and B is minimal, with $n$ of 2.22 and 4.01, respectively. 
However, both areas exhibit path loss values approximately 20 dB higher than  free space model. 
This deviation is attributable to various scatterers present in measurement environment, including low-rise buildings, trees, platforms, utility poles, etc. 
What's more, it can be seen from Fig. \ref{Scenarios} that area A is relatively open with fewer scattering objects, characterized mainly by occasional railway infrastructure and sparse tree cover. 
Consequently,  $n$ for area A approaches   theoretical free-space value of 2. 
In contrast, area B is characterized by a higher density of buildings, contributing to a marginally higher path loss and a correspondingly larger path loss exponent.

Take the result of area A as example, probability density functions (PDFs)  of SF is illustrated in Fig. \ref{PL-SF}(b).
The statistical parameters are listed in Table \ref{summary}.
Standard deviations of areas A and B  are 2.86 dB and 3.40 dB, respectively. 
It can be seen  that shadow fading  can be well fitted to a zero-mean Gaussian distribution. 
Additionally,  standard deviation in area B exhibits greater variability compared to area A, which aligns with the results depicted in Fig. \ref{PL-SF}(a).

\begin{table}[!t]
	\renewcommand{\arraystretch}{1.5}
	\begin{center}
		\caption{Summary of channel characteristic parameters.} 
		\label{summary}
		\begin{tabular}{c| c| c| c c }
			\hline
			\hline
			\multirow{2}{*}{Parameters} & \multirow{2}{*}{\makecell[c]{Statistical  \\ Distribution} } & \multirow{2}{*}{Description} & \multicolumn{2}{c}{Value}    \\
			\cline{4-5}  
			~                   & ~      & ~       &    Area A       &   Area B     \\
			\hline
			PL   & Linear & \makecell[c]{$n$  \\ $A$}  & \makecell[c]{2.22  \\ 49.47} & \makecell[c]{4.01  \\ 9.47} \\
			\hline
			SF   & \makecell[c]{Gaussian  \\ $N(0,2.86^2)$ \\$N(0,3.40^2)$} & \makecell[c]{ Std.}  & \makecell[c]{ 2.86 dB} & \makecell[c]{ 3.40 dB} \\
			\hline
			RMS DS & \makecell[c]{Lognormal  \\ $N(4.33,0.39^2)$} & \makecell[c]{Mean  \\ Std.}  & \multicolumn{2}{c}{\makecell[c]{81.79 ns  \\ 34.39 ns}} \\
			\hline
			\multirow{1}{*}{\makecell[c]{Rice  \\ K-factor}}  & \makecell[c]{Normal  \\ $N(0.66,2.78^2)$ \\$N(-1.22,3.22^2)$}  & \makecell[c]{Mean  \\ Std.}  & \makecell[c]{0.66 dB  \\ 2.78 dB} & \makecell[c]{-1.22 dB  \\ 3.22 dB} \\
			\cline{2-5}
			\hline
			ASA & \makecell[c]{Lognormal  \\ $N(1.78,1.45^2)$} & \makecell[c]{Mean  \\ Std.}  & \multicolumn{2}{c}{\makecell[c]{$16.26^{\circ}$\\$25.19^{\circ}$}} \\
			\hline
			ESA & \makecell[c]{Lognormal  \\ $N(0.48,0.65^2)$} & \makecell[c]{Mean  \\ Std.}  & \multicolumn{2}{c}{\makecell[c]{$2.37^{\circ}$\\$1.91^{\circ}$}} \\
			\hline
			\makecell[c]{Cluster  \\ lifetime}  & \makecell[c]{Lognormal  \\ $N(0.88,0.92^2)$} & \makecell[c]{Mean  \\ Std.}  & \multicolumn{2}{c}{\makecell[c]{3.74 s\\4.57 s}} \\
			\hline
			\makecell[c]{Stationary  \\ Region}  & \makecell[c]{Lognormal  \\ $N(2.16,0.29^2)$} & \makecell[c]{Mean  \\ Std.}  & \multicolumn{2}{c}{\makecell[c]{9.02 m / 0.41 s \\2.51 m / 0.11 s }} \\
			\hline
			\hline
		\end{tabular}
	\end{center}
\end{table}

\subsection{Power Delay Profile and Delay Spread}
\begin{figure}[!t]
	\centering
	\includegraphics[width=.46\textwidth]{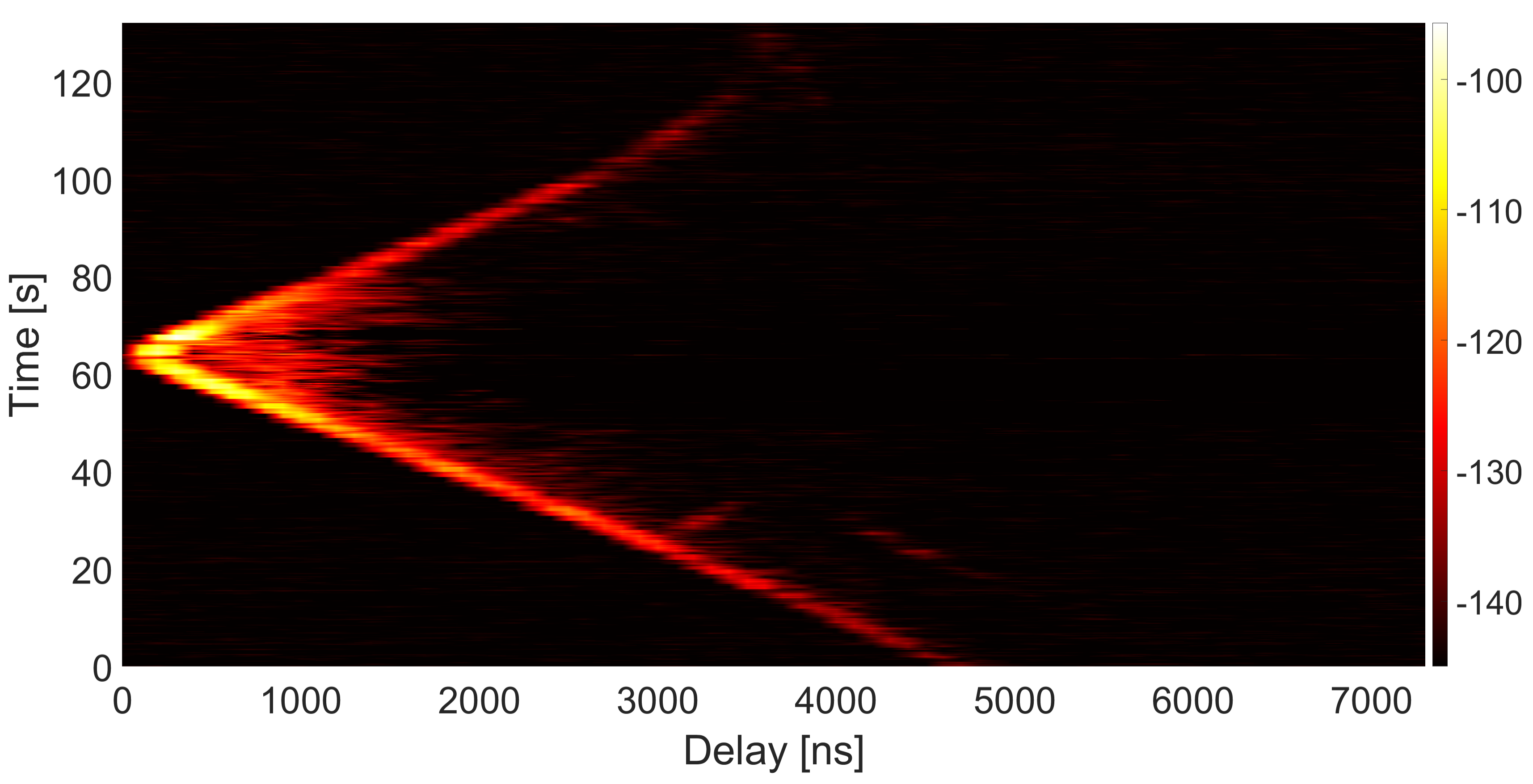}%
	\caption{Measured APDP.}                    
	\label{PDP}
\end{figure}

PDP is extensively employed to characterize the power levels of received paths with propagation delays and to describe the distribution of multi-path components (MPCs) in measured environments.
The instantaneous PDP is denoted as
\begin{equation}
	\label{math_6}
	P(t,\tau ) = {\left| {h(t,\tau )} \right|^2}.
\end{equation}
To achieve more precise analysis results, a fixed noise threshold is used to eliminate noise components, and the noise threshold is set to 6 dB above the background noise floor.
Only signals exceeding this noise threshold are deemed valid, while samples below the threshold are set to zero.
Fig. \ref{PDP} illustrates  average PDPs (APDPs), obtained by averaging with a sliding window of $40\lambda$.

APDPs depicted in Fig. \ref{PDP} aligns closely with  actual measurement scenario. 
As  test train approaches BS and subsequently moves away, the delay of  LOS path correspondingly decreases and then increases. 
The power  of   LOS reaches   peaks when  train is closest to BS. 
Notably, MPCs are prominently observed only for a brief duration near BS. 
This is attributed to the stronger MPCs having more substantial reflection and scattering off objects such as tower, equipment room and trains during this period, 
while in other delay bins, LOS propagation predominates.
This observation can be explained by two factors: firstly,  measurement environment is a rural area with sparse scattering objects and  limited MPCs; secondly, measurement bandwidth is only 10 MHz, resulting in a delay resolution of 100 ns. 
The low delay resolution restricts the ability to distinguish more MPCs.

RMS DS  is the square root of the second central moment of APDPs and is widely used to characterize the delay dispersion of channels. It is defined as
\begin{equation}
	\label{math_7}
	{\tau _{rms}}(d) = \sqrt {\frac{{\sum\nolimits_p {APDP(d,{\tau _p})\tau _p^2} }}{{\sum\nolimits_p {APDP(d,{\tau _p})} }} - {{(\frac{{\sum\nolimits_p {APDP(d,{\tau _p}){\tau _p}} }}{{\sum\nolimits_p {APDP(d,{\tau _p})} }})}^2}}, 
\end{equation}
where $\tau _p$ represents the delay of the $p$th path and $APDP(d,{\tau _p})$ describes the corresponding power with $\tau _p$ measured at location $d$.

\begin{figure}[!t]
	\centering
	\includegraphics[width=.46\textwidth]{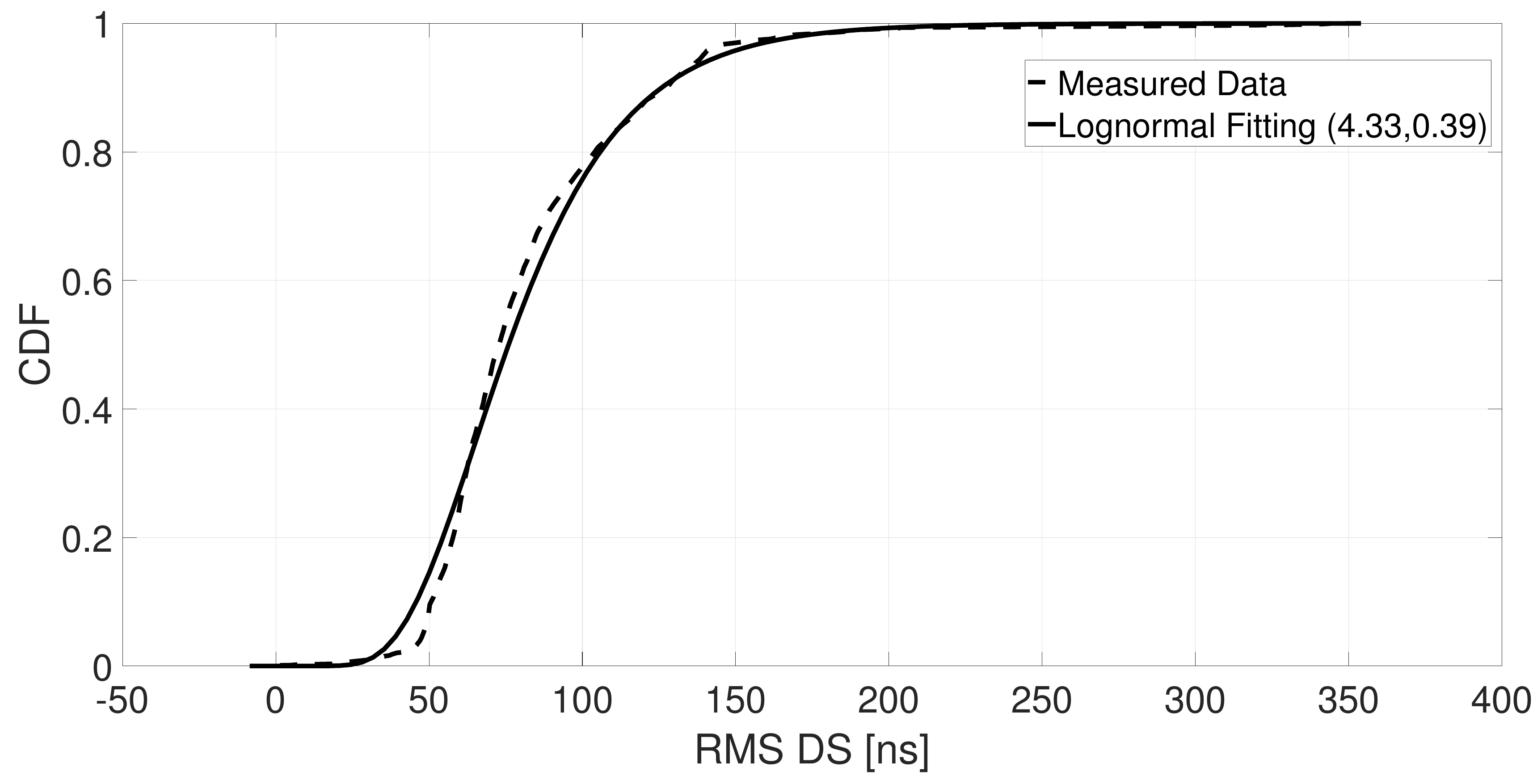}%
	\caption{Measured RMS DS and Lognormal fitting.}                    
	\label{DS}
\end{figure}

The measured results and cumulative probability functions (CDFs) of RMS DS are shown in Fig. \ref{DS}, and statistical results are summarized in Table \ref{summary}.
It can be found that RMS DS  follows a Lognormal distribution, with a mean of 81.79 ns and a standard deviation of 34.39 ns. 
For 90\% of  time, RMS DS remains below 128 ns, while it fluctuates between 150 and 345 ns only for brief periods near BS. 
This phenomenon occurs because proximity to BS results in more abundant MPCs, leading to greater delay spread. 
Conversely, in instances where LOS propagation is predominant,  delay spread is relatively small. 
These findings are highly consistent with those illustrated in Fig. \ref{PDP}.

%

\subsection{Rice K-factor}
Rice K-factor represents the ratio of the power of LOS component and the power of NLOS component in the channel \cite{MiYang2023}, which can be calculated based on APDP.
The calculation formula is as follow
\begin{equation}
	\label{math_8}
	K = \frac{{{{\left| {{h_{LOS}}} \right|}^2}}}{{\sum\limits_\tau  {{{\left| {{h_{NLOS}}} \right|}^2}} }},
\end{equation}
where $h_{LOS}$ is LOS component, and $h_{NLOS}$ is NLOS component. 
All valid multipaths in APDPs are identified using multipath discrimination algorithm \cite{fyy}. 
The multipath with the highest power in single snapshot is designated as  $h_{LOS}$, while the remaining valid multipaths are classified as $h_{NLOS}$.

\begin{figure}[!t]
	\centering
	\subfloat[]{\includegraphics[width=.45\textwidth]{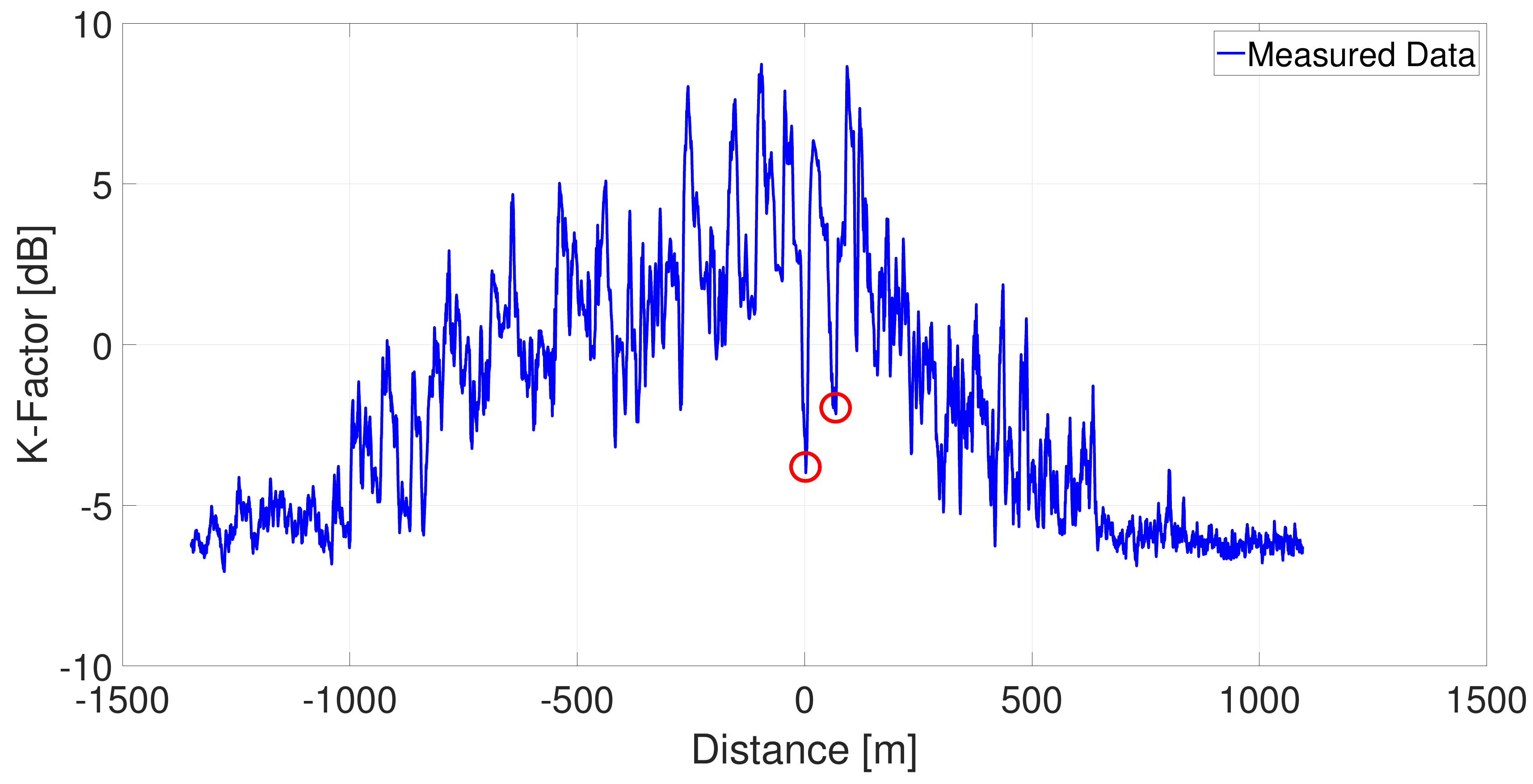}%
		\label{K}}
	\quad
	\subfloat[]{\includegraphics[width=.45\textwidth]{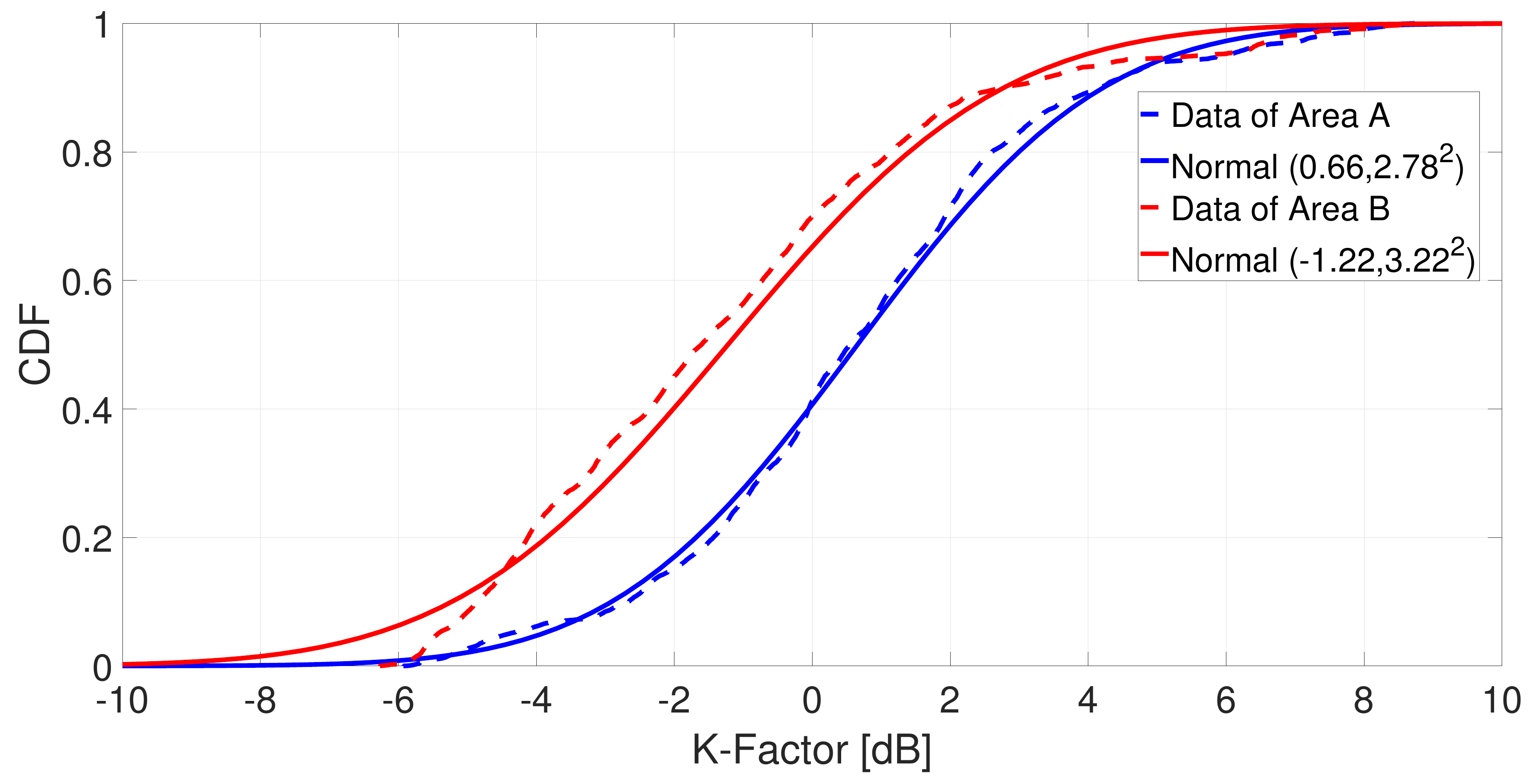}%
		\label{K-CDF}}
	\caption{
		Measured Rice K-factor and corresponding fitting.
		(a) Rice K-factor based on all measured data. The distance equal to 0 represents the location of BS, the range of distance [-1500,0] is area A, and the opposite is area B.
		(b) CDFs of Rice K-factor and Normal fitting of  areas A and B.
	}
	\label{K-Factor}
\end{figure}

Rice K-factor derived from  full-range measured data is depicted in Fig. \ref{K-Factor}(a). 
BS is located at the origin (distance equals zero), Area A spans from -1500 to 0 m, and Area B extends from 0 to 1500 m. 
Fig. \ref{K-Factor}(a) indicates that  K-factor generally increases as the distance decreases, implying a higher proportion of power from LOS.
However, a sharp decline is observed directly below BS as shown  the red circles, attributed to   near-shadow area caused by antenna  directivity.
Figs. \ref{K-Factor}(b) presents the corresponding CDFs and  Normal distribution fitting of   K-factor in two areas. 
It can be seen  that    K-factor follows a Normal distribution, with mean and standard deviation values in areas A and B being 0.66, 2.78 dB and -1.22, 3.22 dB, respectively. 
The smaller mean and larger standard deviation in area B imply a greater proportion of NLOS components and more significant fluctuations, consistent with the observations in Fig. \ref{K-Factor}(a).

\subsection{Stationary Region}
\begin{figure}[!t]
	\centering
	\subfloat[]{\includegraphics[width=.25\textwidth]{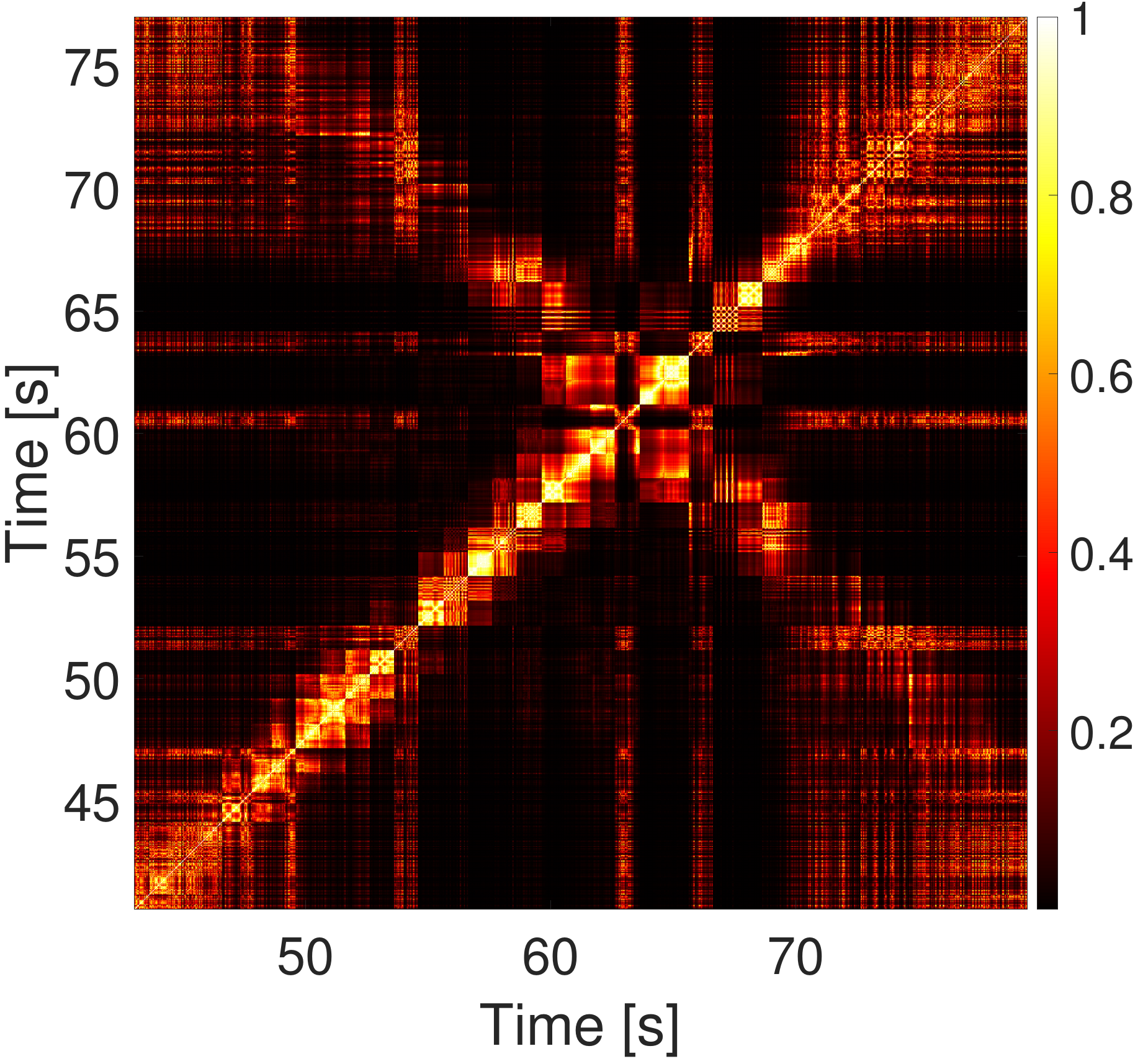}%
		\label{TPCC}}
	\subfloat[]{\includegraphics[width=.235\textwidth]{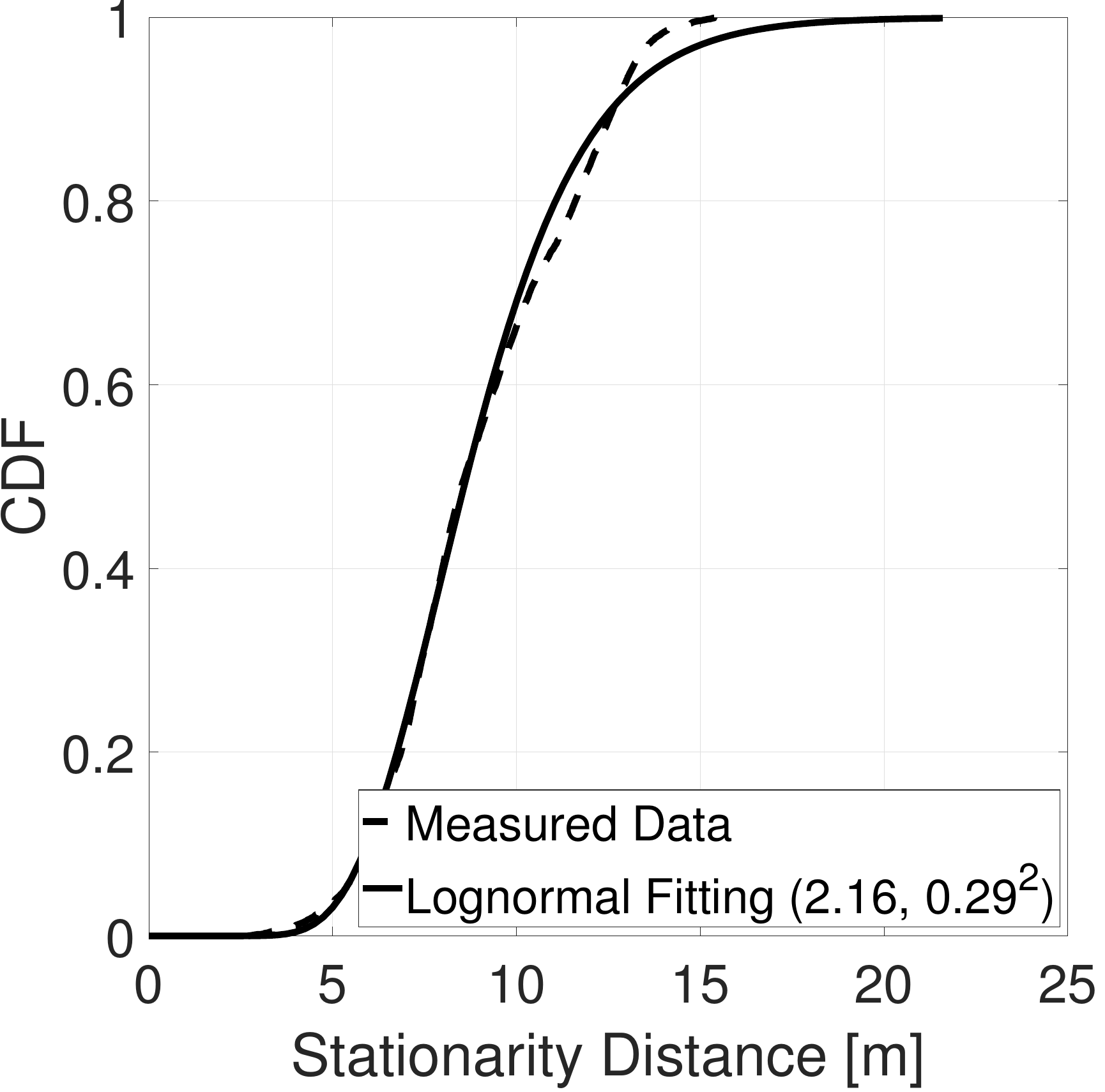}%
				\label{stationary_CDF}}
	\caption{
			TPCCs and stationarity distance.
			(a) TPCCs;
			(b) CDFs of stationarity distance.
		}
	\label{Stationary}
\end{figure}

The rapid movement of high-speed train    causes channel non-stationarity.
In this paper, the temporal PDPs correlation coefficient (TPCC) is used as a measure of channel stationarity. 
The temporal similarity of PDPs between different times can be quantified by this metric.
The TPCC between the PDPs at time $t_i$ and $t_j$ can be computed as
\begin{equation}
	\label{math_88}
	{c\left( {{t_i},{t_j}} \right) = \frac{{\int {P\left( {{t_i},\tau } \right)}  \cdot P\left( {{t_j},\tau } \right)d\tau }}{{\max \left\{ {\int {{P^2}\left( {{t_i},\tau } \right)} d\tau ,\int {{P^2}\left( {{t_j},\tau } \right)} d\tau } \right\}}}.}
\end{equation}
The values of TPCCs are normalized from 0 to 1,
and a high TPCC value indicates that the channel has higher similarity between $t_i$ and $t_j$, as shown in Fig. \ref{Stationary}(a). 
What's more, a threshold needs to be selected to determine stationary region window $\Delta W$. 
Following the recommendation of \cite{ym2020}, 0.8 is chosen as  threshold.
For a TPCC between  $t_i$ and $t_j$, when $c\left( {{t_i},{t_j}} \right)$ is large than the threshold, the channel is considered not to experience significant change. 
Then $\Delta W$ between  $t_i$ and $t_j$  can be obtained from TPCCs by using the threshold.
Further, we  calculates the stationarity distance  by using the relationship between time, distance, and velocity. 
Fig. \ref{Stationary}(b) statistics the CDFs of stationarity distance for measured data and Lognormal fitting. 
The average values and standard deviation of stationarity distance  are 9.02 m and 2.51 m  respectively, and the corresponding $\Delta W$ is 0.41 s and 0.11  s.

\section{Multipath Cluster Characteristics}

In this section, we extract and analyze  multipath clusters characteristics based on multi-antenna data, focusing on estimating  MPC parameters using the Space-Alternating Generalized Expectation-Maximization (SAGE) algorithm and identifying and tracking  clusters with KPowerMeans and multipath component distance (MCD)-based tracking algorithms.
It is important to note that at greater distances from BS, the signal-to-noise ratio   of  multi-antenna receiver is lower compared to that of  single-antenna receiver. 
As a result, although measured simultaneously, effective measurement duration for  multi-antenna data are shorter, approximately 65 s and 1000 snapshots, while  single-antenna data extends to around 123 s.


\begin{figure*}[!t]
	\centering
	\subfloat[]{\includegraphics[width=.46\textwidth]{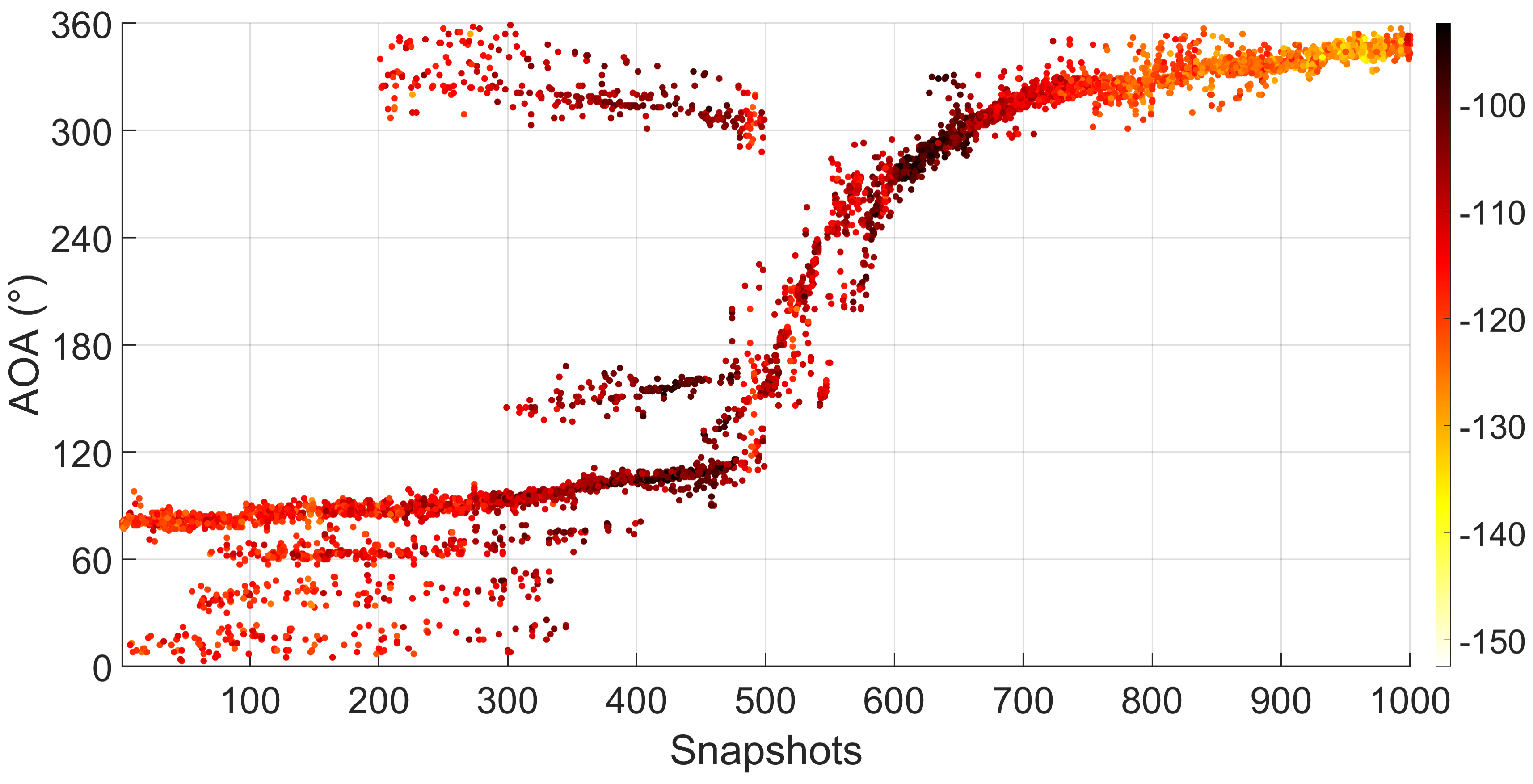}%
		\label{AOA}}
	\subfloat[]{\includegraphics[width=.46\textwidth]{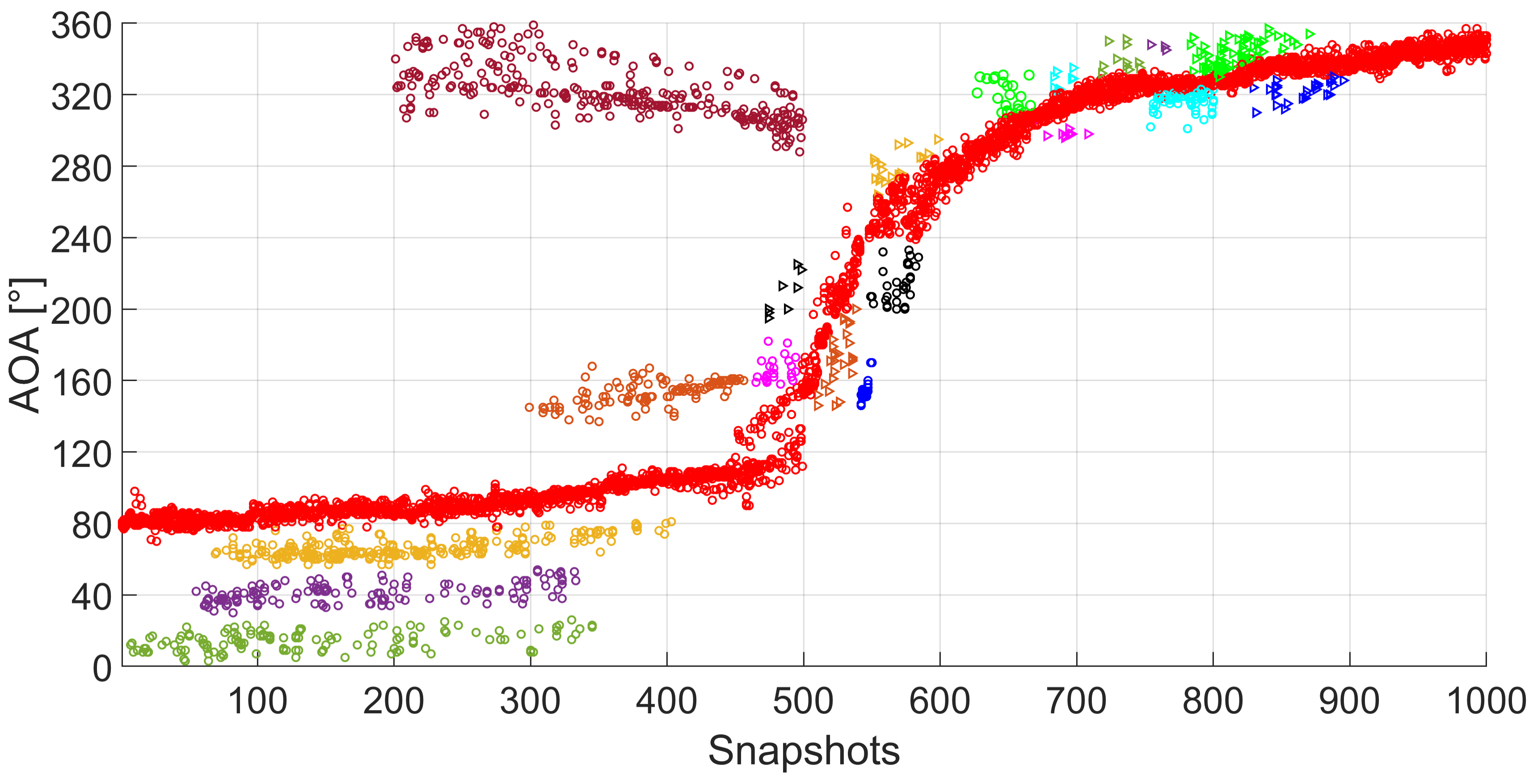}%
		\label{cluster}}
	\caption{
		(a) Angle estimation results of AOA. 
		(b) Cluster identification and tracking results.
	}
	\label{Angular}
\end{figure*}

\subsection{Estimation of MPCs Parameters}
As the antenna array is only utilized at Rx and not at Tx,
we are limited to obtaining arrival angle information.
It is important to note that we utilize a circular array, with AOA  and EOA calculated based on  steering vector within the ranges of [$0^{\circ}$, $360^{\circ}$] and [$0^{\circ}$, $90^{\circ}$], respectively. 
The  SAGE algorithm is employed to extract MPC parameters, including power, delay, AOA and EOA, i.e. $ \left\{ {\alpha ,\tau ,\phi ,\theta } \right\}  $.
The SAGE algorithm provides a maximum-likelihood estimate of the MPC parameters through an iterative process, which has been wildly applied to MPC identification and parameter estimation. 
Then the angular resolution of  MPC angle estimation result obtained by  SAGE algorithm in this paper is $2^{\circ}$.
Detailed description of SAGE algorithm can be found in \cite{yangmi2019}.

The amplitude  and angle of arrival of MPCs derived from measured data are illustrated in Fig. \ref{Angular}(a).
It is observed that as Rx gradually approaches and then moves away from BS, AOA of LOS component changes from $80^{\circ}$ to $360^{\circ}$, consistent with the findings in \cite{zhoutao2019}. Notably, around the 450-th to 600-th snapshots, Rx passes directly beneath  BS, causing  AOA to become discretized due to the presence of  near-shadow area. 
Additionally, several distinct MPC clusters, resulting from large buildings adjacent to the railway in the measurement environment, are observed besides  LOS component.
Conversely,  EOA generally exhibits minor fluctuations around $50^{\circ}$ to $80^{\circ}$, except for a brief peak when approaching and moving away from  BS.
The fluctuations EOA are not demonstrated here, but can be seen in previous work \cite{zxjwcsp}.

RMS AS is defined as the second moment of  angular power spectrum  and is typically used to describe  angular dispersion of  channel. 
It is widely used to represent the discreteness of arrival/departure angles, 
and can be calculated as
\begin{equation}
	\label{math_10}
	AS = \sqrt {\sum\limits_{l = 1}^L {{{({\Psi _{Angle,l}} - {\mu _{APS}})}^2}\alpha _l^2/\sum\limits_{l = 1}^L {\alpha _l^2} } },
\end{equation}
where $L$ is the total number of all MPCs. 
Parameters $\Psi _{Angle,l}$ and $\alpha _l$ represent   arrival angle and complex amplitude of the $l$th MPC, respectively. 
Parameter $\mu _{APS}$ is the mean of angular power spectrum and
\begin{equation}
	\label{math_11}
	{\mu _{APS}} = \sum\limits_{l = 1}^L {{\Psi _{Angle,l}}\alpha _l^2} /\sum\limits_{l = 1}^L {\alpha _l^2}.  
\end{equation}
Statistical parameters of ASA and ESA are listed in Table \ref{summary}.
ASA and ESA generally conform to  Lognormal distribution, with  means and standard deviations being $16.26^{\circ}$, $25.19^{\circ}$ and $2.37^{\circ}$, $1.91^{\circ}$, respectively. 
Among them, the mean and standard deviation of ASA are much larger than those of ESA, which indicates that diffusion degree of AOA is greater.

\subsection{Identification and Tracking of MPCs Clusters}
Based on  MPCs parameter estimations, clustering and tracking algorithm are used to identify time-variant clusters of MPCs. 
As an evolved algorithm of KMeans and incorporating the MPC power as the weight \cite{czz2022}, classical KPowerMeans algorithm is   employed to cluster MPCs in this paper.
A detailed describption of  KPowerMeans algorithm can be found in \cite{kpm}.
Subsequently, we apply a  MCD-based tracking algorithm to identify and track time-variant MPC clusters across multiple snapshots. 
For two arbitrary MPCs $ {P_x}  $ and $ {P_y}  $ in consecutive snapshots  $ {\Omega _i}  $ and $ {\Omega _{i+1}}  $,  the MPCs’ information can be expressed as  \cite{he2015,ljz2019}
\begin{equation}
	\label{math_12}
	{\begin{array}{l}
			{P_x} \in {\Omega _i}:\left[ {{\alpha _x}\left( i \right);{\tau _x}\left( i \right);{\phi _x}\left( i \right);{\theta _x}\left( i \right)} \right],\\
			{P_y} \in {\Omega _{i + 1}}:\left[ {{\alpha _x}\left( {i + 1} \right);{\tau _x}\left( {i + 1} \right);{\phi _x}\left( {i + 1} \right);{\theta _x}\left( {i + 1} \right)} \right].
		\end{array}}  
\end{equation}
The MCD of  two MPCs in  angle domain can be expressed as
\begin{equation}
	\label{math_13}
	{\begin{array}{l}
			MC{D_{angle,xy}}\\
			= \frac{1}{2}\left| {\left( {\begin{array}{*{20}{c}}
						{\cos \left( {{\phi _x}} \right)\sin \left( {{\theta _x}} \right)}\\
						{\sin \left( {{\phi _x}} \right)\sin \left( {{\theta _x}} \right)}\\
						{\cos \left( {{\theta _x}} \right)}
				\end{array}} \right) - \left( {\begin{array}{*{20}{c}}
						{\cos \left( {{\phi _y}} \right)\sin \left( {{\theta _y}} \right)}\\
						{\sin \left( {{\phi _y}} \right)\sin \left( {{\theta _y}} \right)}\\
						{\cos \left( {{\theta _y}} \right)}
				\end{array}} \right)} \right|.
		\end{array}}  
\end{equation}
The MCD of delay is obtained as
\begin{equation}
	\label{math_14}
	{MC{D_{delay,xy}} = \xi  \cdot \frac{{{\tau _{std}}}}{{\Delta {\tau _{\max }}}} \cdot \frac{{\left| {{\tau _x} - {\tau _y}} \right|}}{{\Delta {\tau _{\max }}}},}  
\end{equation}
where
\begin{equation}
	\label{math_15}
	{\Delta {\tau _{\max }} = \max \left( {{\tau _n}} \right) - \min \left( {{\tau _n}} \right),n \in \left( {{\Omega _i} \cup {\Omega _{i + 1}}} \right)}  
\end{equation}
and $ \tau _{std}  $ is a standard deviation for the delays of $ {{\Omega _i} \cup {\Omega _{i + 1}}} $. 
Parameter $ \xi  $ is scaling factor used to adjust the weight of $ MC{D_{delay,xy}}  $ in the overall MCD $ MC{D_{xy}}  $.
And $ MC{D_{xy}}  $ can be expressed as follow
\begin{equation}
	\label{math_16}
	{MC{D_{xy}} = \sqrt {{{\left\| {MC{D_{angle,xy}}} \right\|}^2} + MC{D_{delay,xy}}^2} .}  
\end{equation}
MCD measures the similarity between two MPCs, i.e.  
 smaller value means that two MPCS are more similar.
What's more, a specified threshold  is used to judge whether the closest old MPC/cluster can be treated as a new MPC/cluster. 
This threshold is mainly determined by the resolution of measurement system in the angular and delay domain, as well as the data post-processing scheme. 
The value of  threshold is set to 0.06 by experience and is found to produce satisfying tracking results.
To address the limitations of automatic algorithms in accurately capturing small scatterer clusters, we further refine the clustering results through visual inspection. 
This approach effectively balances accuracy and computational efficiency, and is also used in \cite{ljz2018, ym2020}.

Fig. \ref{Angular}(b) demonstrates the result of  clustering and tracking  based on measured data,  where each cluster is distinguished by a unique color or marker. 
Specifically,  red circles denote LOS cluster, while the remaining markers represent scatterer clusters.
It can be seen   that the LOS cluster  is dominant and consistently present. 
However, limited by  measurement scenario and  bandwidth, although a few long-lasting and prominent clusters can be identified, the total number of distinguishable clusters is limited.

\subsection{Time-Variant Evolution Characterization}
The high mobility of high-speed trains will lead to rapid time-variant channels, which in turn leads to  birth and death of clusters and rapid changes in cluster characteristics \cite{zxw2022}.
In Section II, we present  a general model of   non-stationarity in 5G-R channels and the birth-death process of clusters. 
In this subsection,  we utilize cluster lifetime and a first-order four-state Markov chain to characterize  birth-death process of clusters based on  measured data.

\begin{figure}[!t]
	\centering
	\includegraphics[width=.46\textwidth]{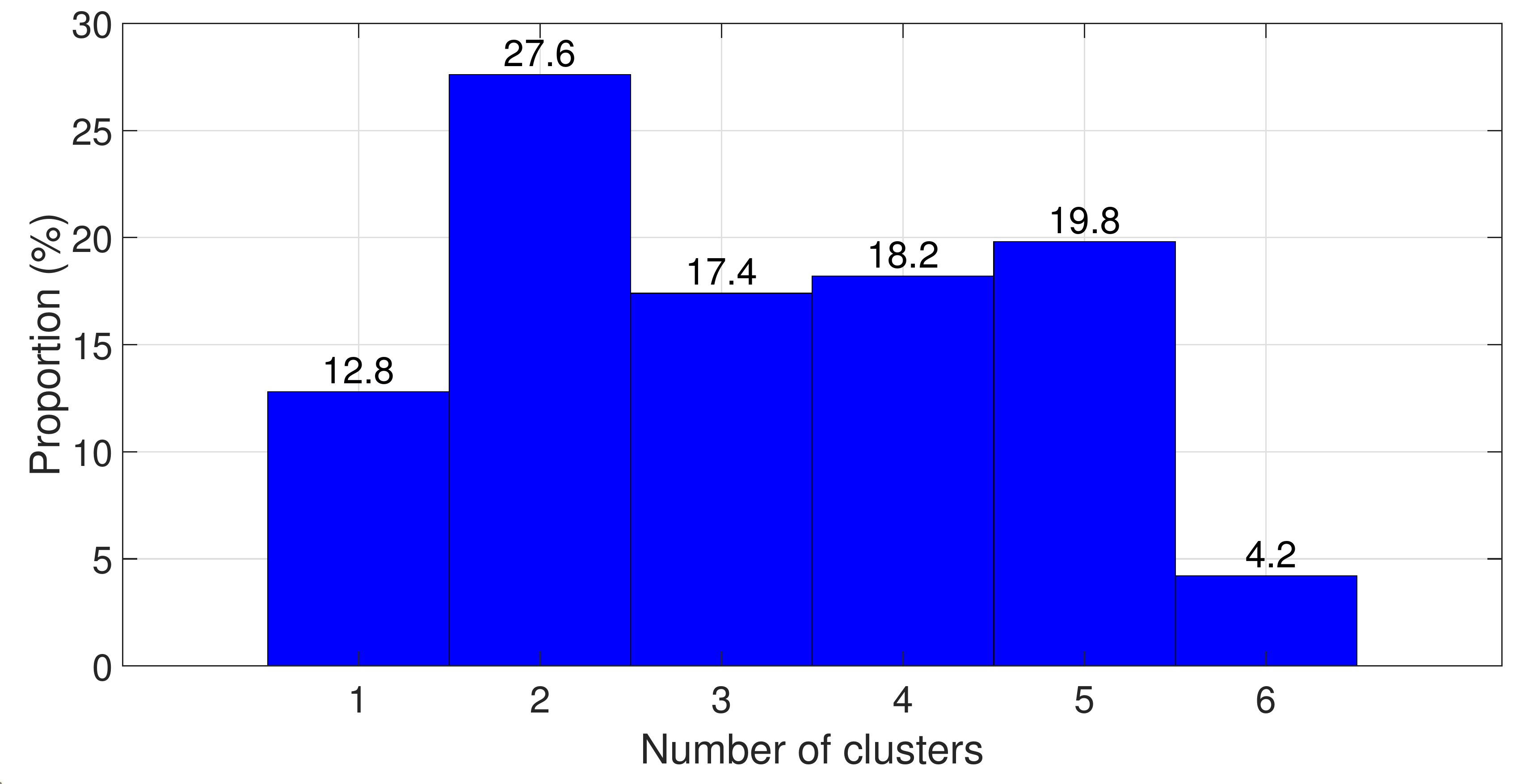}%
	\caption{The count histogram of the cluster number within stationary region $\Delta W$.}                    
	\label{numclu}
\end{figure}

\textit{1) Number of Clusters.}
The number of clusters can reflect the richness of MPCs in measurement environment \cite{ym2019}.
We  have identified a total of 20 clusters within an effective measurement period of approximately 60 seconds, including one LOS cluster and 19 NLOS clusters, as shown in Fig. \ref{Angular}(b). 
Fig. \ref{numclu} illustrates the proportion of clusters observed within  stationary region $\Delta W$. 
It can be seen that in roughly one-third of the time, only two clusters are distinguishable, while six clusters can be identified in just 4\% of the time. 
Additionally, five clusters are observable in about one-fifth of  time,  which is slightly higher than 3 and 4 clusters.
This phenomenon can be attributed to two factors. 
First, the relatively open measurement environment means that only certain houses or buildings along the railway act as scatterers and has low scatterer density. 
Second,  small measurement bandwidth results in low delay resolution, making it difficult to distinguish more MPCs.

\begin{figure}[!t]
	\centering
	\includegraphics[width=.46\textwidth]{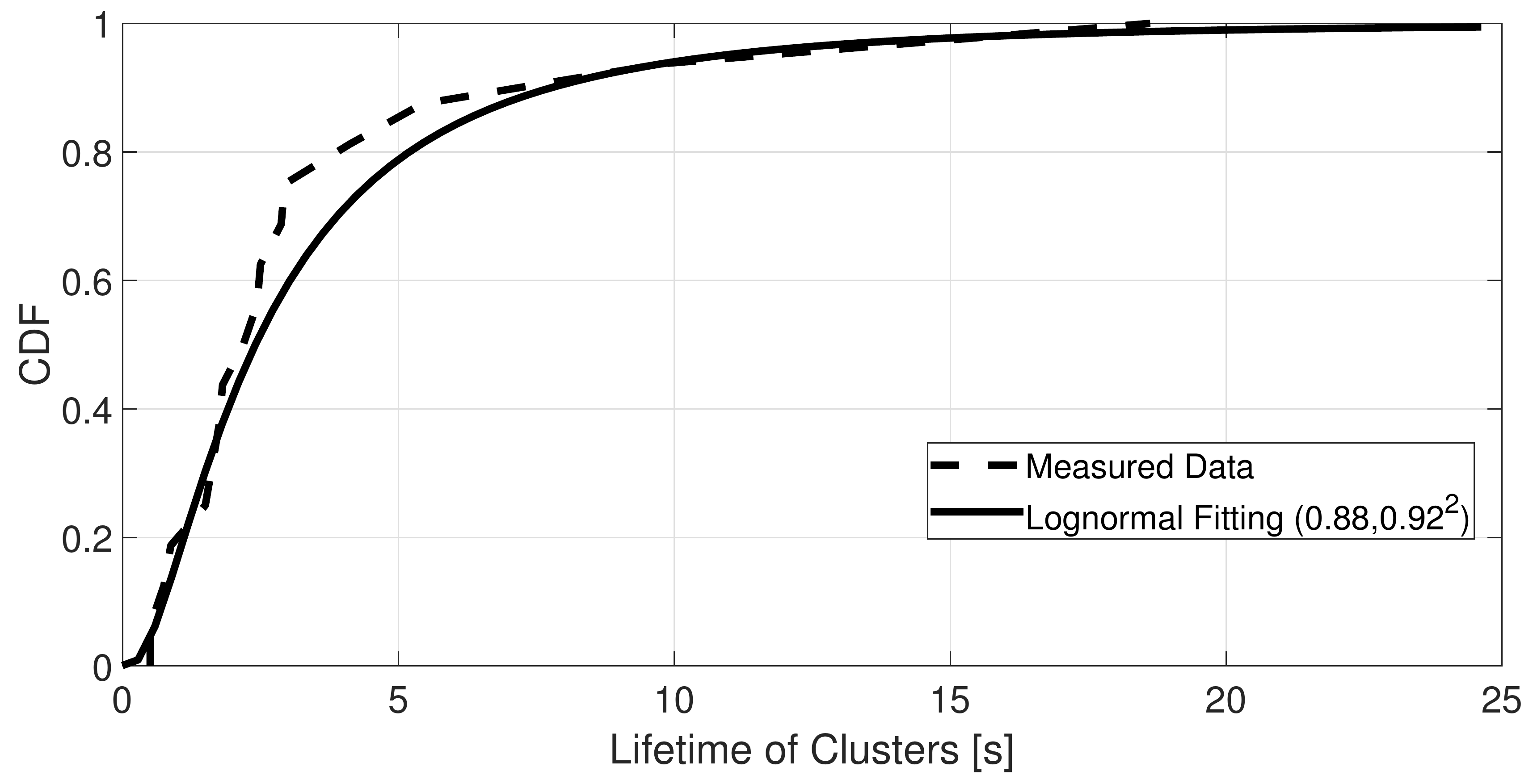}%
	\caption{The CDFs of cluster lifetime.}                    
	\label{lifetime}
\end{figure}


\begin{figure}[!t]
	\centering
	\includegraphics[width=.3\textwidth]{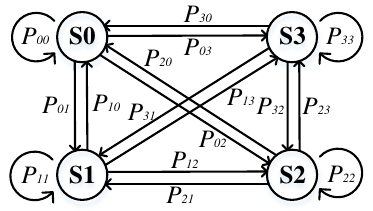}%
	\caption{First-order four-state Markov chain.}                    
	\label{markovchain}
\end{figure}

\textit{2) Lifetime of Clusters.}
The lifetime of clusters is the duration between the appearance and disappearance of each cluster.
Fig. \ref{lifetime} shows the CDFs of  cluster lifetime. 
It is found that cluster lifetime   has a Lognormal distribution, with  mean value 3.74 s and standard deviation 4.57 s.
It is seen that 80\% of  clusters have a lifetime of less than 5 seconds, indicating that most scatterers are relatively small in size. 
However, a few clusters exhibit longer lifetimes, ranging from 15 to 20 seconds. 
This is due to the presence of larger structures, such as control rooms and operation and maintenance facilities along the railway,  which can be observed from Fig. \ref{Scenarios}.

\textit{3) Markov chain.}
Markov chains are often used to characterize the evolution of dynamic clusters \cite{zzl2024,zt2022}.
In order to analyze  birth-death process of dynamic clusters, a first-order four-state Markov chain is considered, as shown in Fig. \ref{markovchain}.
Four typical states conform to the characteristics of 5G-R communication channel evolution.
The four states are defined as $ {S_0}  $ (no ``births" or ``deaths"), $ {S_1}  $ (``births" only), $ {S_2}  $ (``deaths" only), and $ {S_3}  $ (both ``births" and ``deaths").
Based on clustering rusults,
state transfer matrix $ {P_T}  $  is 
\begin{equation}
	\label{math_17}
	{\begin{array}{l}
			{P_T}  = \left\{ {{p_{ij}}} \right\} = \left[ {\begin{array}{*{20}{c}}
					{{P_{00}}}&{{P_{01}}}&{{P_{02}}}&{{P_{03}}}\\
					{{P_{10}}}&{{P_{11}}}&{{P_{12}}}&{{P_{13}}}\\
					{{P_{20}}}&{{P_{21}}}&{{P_{22}}}&{{P_{23}}}\\
					{{P_{30}}}&{{P_{31}}}&{{P_{32}}}&{{P_{33}}}
			\end{array}} \right] \\
			\quad \ \,  = \left[ {\begin{array}{*{20}{c}}
					{0.66}&{0.16}&{0.12}&{0.06}\\
					{0.28}&{0.02}&{0.53}&{0.17}\\
					{0.36}&{0.47}&{0.05}&{0.12}\\
					{0.16}&{0.13}&{0.19}&{0.52}
			\end{array}} \right]   
		\end{array}}  
\end{equation}
where $ i $ and $ j $ represent the state index, while $p_{ij} $ is the transition probability from state $S_{i} $ to state $S_{j} $.
It can be observed that for   initial state  $ {S_0}  $, its next state is most likely to remain at  $ {S_0}  $ with a high probability, as indicated by the largest transition probability  $ P_{00} $. 
It implies that, in the absence of cluster birth or death, next state will most likely continue without any cluster birth or death, corresponding to the first half of Fig. \ref{Angular}, where only LOS cluster and a few long-lasting NLOS clusters are present.
For initial states $ {S_1}  $, $ {S_2}  $ and $ {S_3}  $, there is approximately a 50\% chance that they will transition sequentially to states $ {S_2}  $, $ {S_1}  $ and $ {S_3}  $  respectively, 
which means  that when cluster birth or death occurs, the subsequent state is highly likely to   be the opposite state or remain unchanged. 
This indicates a relatively high density of clusters, with a more uniform spatial distribution, 
which   aligns with the second half of Fig. \ref{Angular},  reflecting the presence of numerous and short-duration scatterers.
The above phenomenon can also be clearly observed in  measurement scenario as shown in  Fig. \ref{Scenarios}.  
There are few low-rise buildings such as houses in area A, but larger railway-specific structures dominate the environment, whereas in area B, the presence of denser housing results in the frequent birth and death of clusters.

\begin{table*}
	\renewcommand{\arraystretch}{1.5} 
	\begin{center}
		\caption{CDL Model Parameters.}
		\label{CDL}
		\begin{threeparttable}
			\begin{tabular}{ c| c| c| c| c| c|  c| c| c }
				\hline
				\hline
				Model & \multicolumn{4}{c|}{5G-R Rural} & \multicolumn{4}{c}{3GPP RMa CDL-D \cite{3gpp}}   \\
				\hline
				No.  & Delay [ns]  & Power [dB] & AOA [deg] & EOA [deg]    &  Delay [ns] & Power [dB] & AOA [deg] & EOA [deg]  \\
				\hline
				\makecell[c]{1 \\(LOS)}   & 0 & -0.5 & 219.6 & 65.5 &  0 & -0.2 & -180 & 81.5   \\
				\hline
				2    & 70.787 & -23.7 & 153.5 & 70.9  & 5.497 & -18.8 & 89.2 & 86.9   \\
				\hline
				3    & 180.345 & -20.9 & 166.6 & 66.8  & 96.123 & -21 & 89.2 & 86.9   \\
				\hline
				4    & 282.813 & -11.4 & 153.5 & 65.2  & 214.08 & -22.8 & 89.2 & 86.9   \\
				\hline
				5    & 806.152 & -15.1 & 66.2 & 64.5  & 220.674 & -17.9 & 163 & 79.4   \\
				
				\hline
				\hline
				
			\end{tabular}
			
		\end{threeparttable} 
	\end{center}
\end{table*}

\section{MODEL IMPLEMENTATION AND VALIDATION}
\subsection{Measurement-Based CDL Channel Model}
In Sections IV and V, we have obtained sufficient channel characteristics and MPC cluster characteristics. 
Next, we  establish the CDL model of 5G-R channel according to  standard procedure of 3GPP TR 38.901 \cite{3gpp}.
Five CDL models are mentioned in \cite{3gpp},
and the widely used CDL-D model in \cite{czz2022, rural2020} for LOS scenarios are
selected  as the standard.
There are 13 clusters are defined in CDL-D model, but  such a large number of clusters cannot be obtained based on measurements. 
As illustrated in Fig. \ref{numclu}, within stationary region, 
we can identify a maximum of 6 clusters with  the  lowest proportion. 
While the 2 clusters have the highest proportion, it is not sufficient for developing a generalized CDL model. 
Therefore,  we  choose the scenario with 5 clusters, which represents the second most frequent case, to construct CDL model.
Following the procedure in Section II, we merge  the identified clusters and calculate  the delays, powers, AOAs and EOAs, in accordance with the CDL-D model. 
Since the 3GPP RMa scenario is similar to  test environment, we herein incorporate its CDL model as a comparison.
The parameters of  established CDL model and  the first five clusters in CDL-D model are presented in  Table \ref{CDL}.

\begin{figure*}[!t]
	\centering
	\subfloat[]{\includegraphics[width=.43\textwidth]{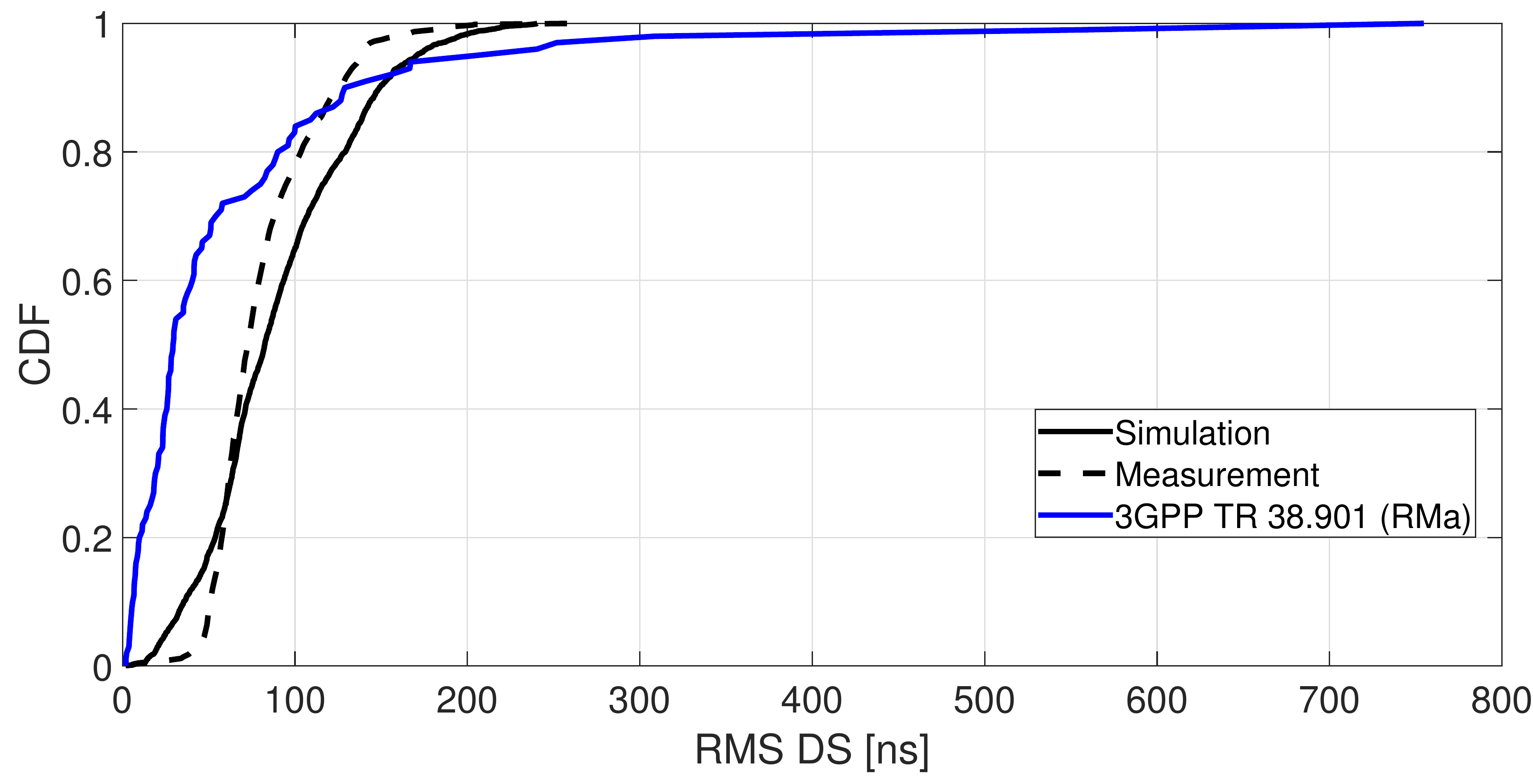}%
		\label{DS_V}}
	\subfloat[]{\includegraphics[width=.43\textwidth]{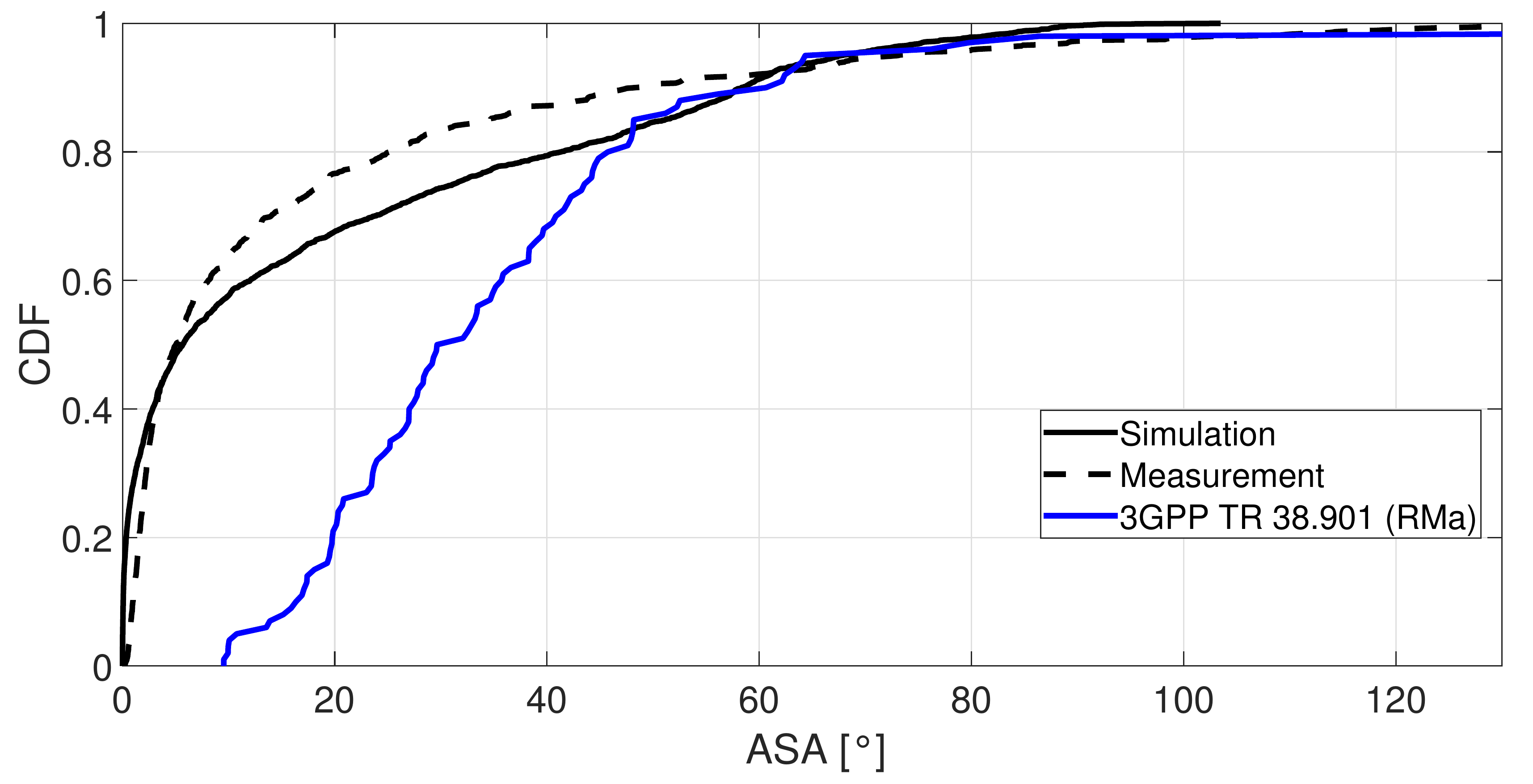}%
		\label{ASA_V}}
	\quad
	\subfloat[]{\includegraphics[width=.43\textwidth]{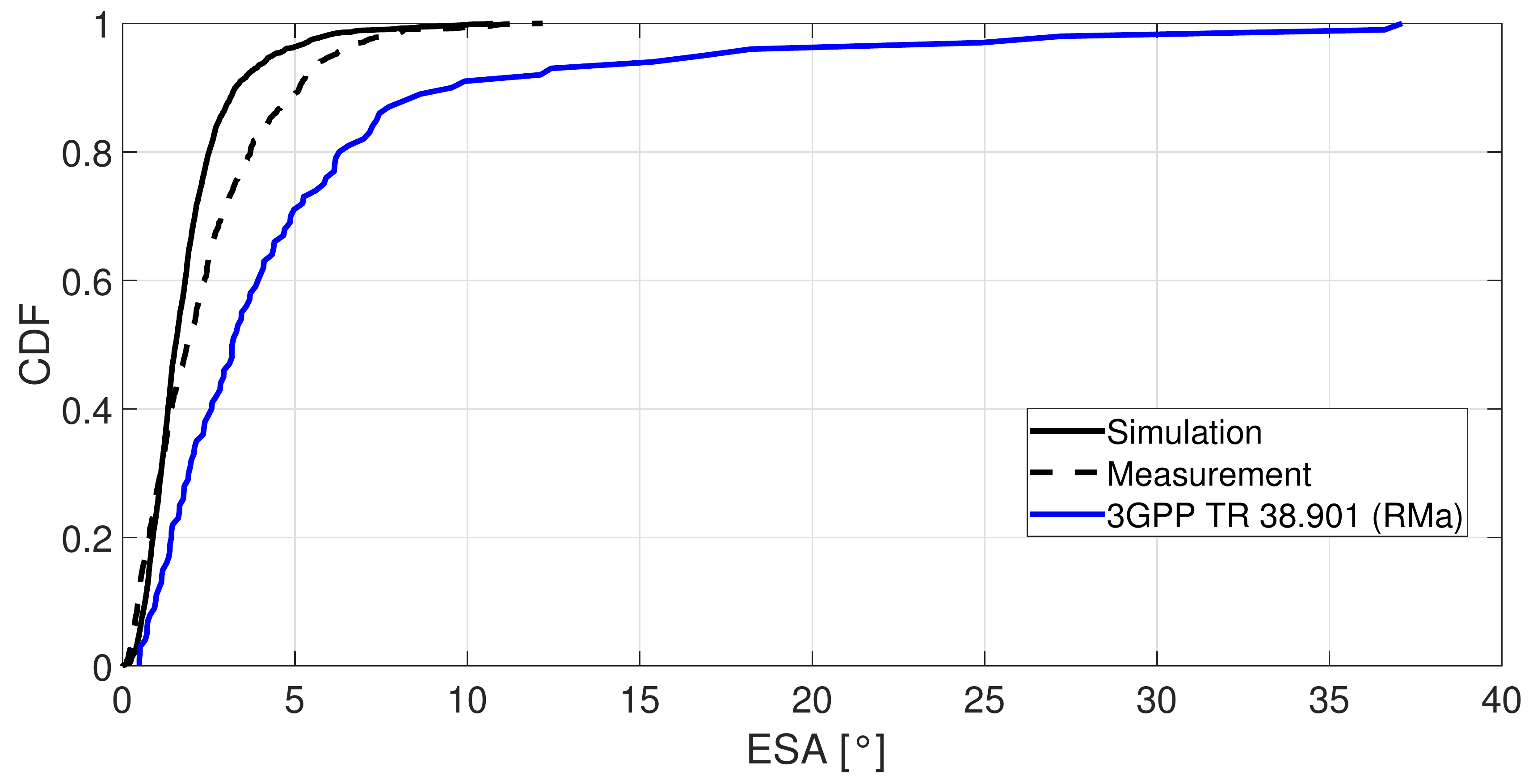}%
		\label{ESA_V}}
	\subfloat[]{\includegraphics[width=.43\textwidth]{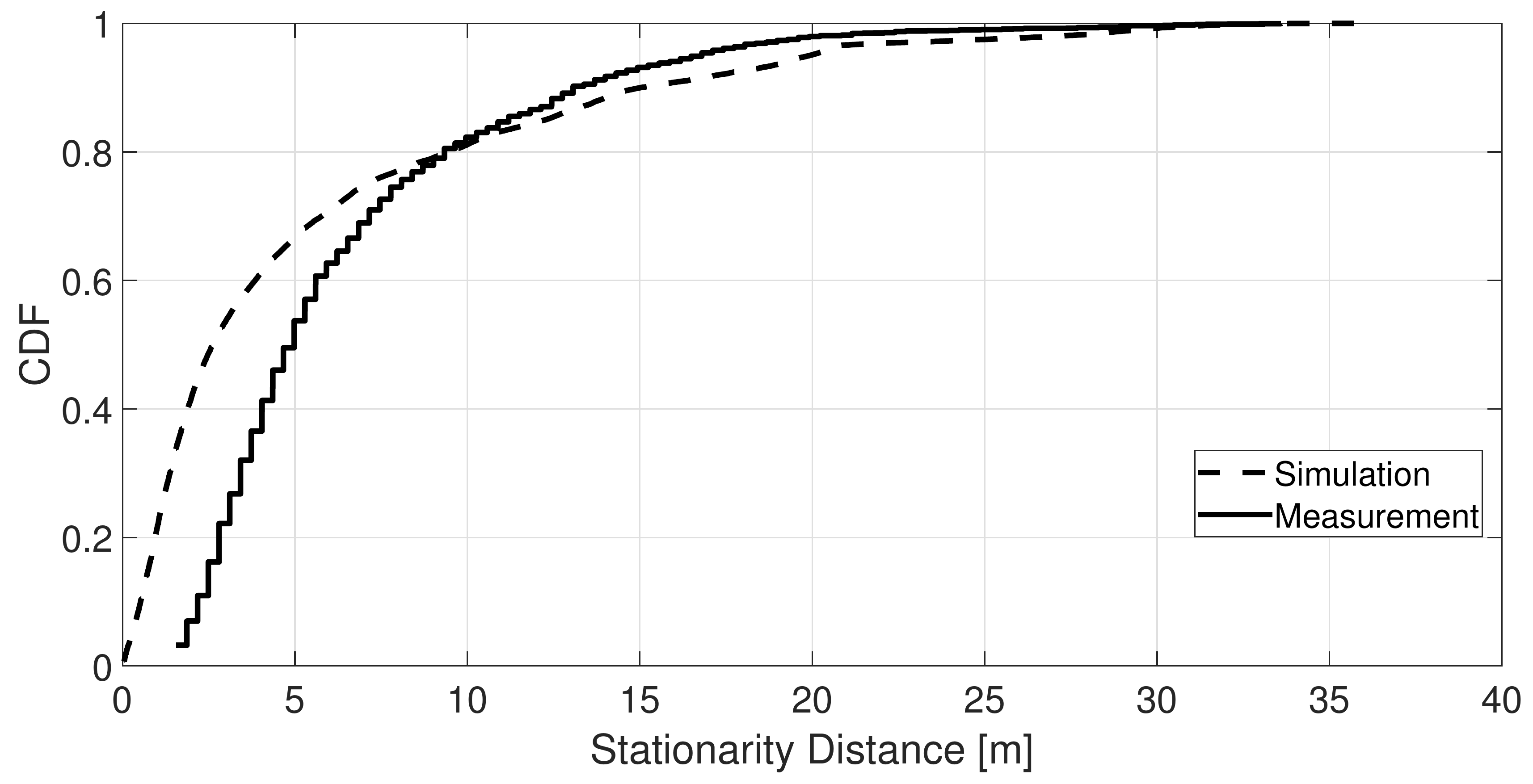}%
		\label{TPCC_V}}
	\caption{
		Model comparisons between simulated and measured channels. (a) RMS DS. (b) ASA. (c) ESA. (d) Stationarity distance.
	}
	\label{validate}
\end{figure*}

For comparison, we scale   normalized delay in   CDL-D model by applying a scaling factor for  RMa scenario to obtain scaled delay, while other parameters are calculated following 3GPP.
We find that  the proposed 5G-R CDL model has several differences from  3GPP CDL-D model. 
First, from the perspective of delay,  cluster delay in  5G-R model is significantly larger than  3GPP model, suggesting that the measurement scenario involves more distant scatterers. 
In addition to the difference in measurement scenarios, this phenomenon may also be be attributed to the limited delay resolution, which is  only 100 nanoseconds,  making it difficult to fully distinguish LOS path and  MPCs close to it. 
Therefore,  MPCs  that can be distinguished are far away from LOS path and have a substantially larger relative delay.
As for power, it can be found that LOS cluster in   5G-R model has a lower power value compared to  3GPP model, indicating that more power is allocated to NLOS clusters.
It can be explained in Fig. \ref{Angular}(b), which shows that when  LOS cluster is positioned directly below BS, its energy is more dispersed, and a portion of  energy is distributed to NLOS clusters. 
This may be the reason why the power of  4th and 5th clusters is still high even though the delay is large.
AOA and EOA are  highly scenario-dependent. 
Since both models are rural scenarios, their overall trends of angles are consistent, though there are slight variations in specific values. 
For example, values of EOA  fluctuate between $64^{\circ}-71^{\circ}$  and $79^{\circ}-87^{\circ}$ in both models, respectively.
This indicates that in  rural scenario, EOA changes remain relatively stable, while  AOA, in contrast, shows more significant fluctuations, even spanning up to $180^{\circ}$.

\subsection{Model Validation}
In this paper, three second-order statistics, i.e.,  RMS DS, ASA and ESA, and stationarity distance are used to validate the proposed 5G-R time-variant model. 
To compare with measured data, we set antenna height, positions and other general simulation parameters to be the same as in measurements.
The generation of channel coefficients follows the process in Section II.
Note that  multiple sets of data are obtained during   measurements. 
Part of the dataset is used for statistical analysis and modeling, while the rest is used for model validation.

The CDFs of RMS DS, AS, stationarity distance between simulated, measured and 3GPP channels are shown in Fig. \ref{validate}, respectively. 
Note that  since  3GPP   lacks parameterization for describing non-stationarity, a comparison with   3GPP  model in Fig. \ref{validate}(d) is not possible. 
However,   3GPP model is included for comparison alongside simulations and measurements  in the other three subfigures.
Due to the positions of scatterers in   simulation are randomly generated, the resulting channel parameters cannot fully align with the measurements, even though the UT parameters are configured to match the measurements as closely as possible. 
Nevertheless, the overall agreement is fairly good and reasonable.
In contrast, the fitting results of   3GPP channel and  measured channel are quite different.
This clearly shows that the proposed model   has improved performances than  3GPP model, making it more consistent with  actual dynamic channel conditions.

\section{CONCLUSION}
In this paper, we present a   cluster-based  time-variant channel model for 5G-R within an enhanced 3GPP framework, incorporating time evolution to capture   dynamic channel characteristics.
Extensive channel measurements are conducted on  5G-R private network test line of China, providing a rich measured data for analysis. 
We then   analyze typical channel fading characteristics, such as PL, RMS DS, AOA, etc. 
Additionally, we extract MPCs  using   KPowerMeans algorithm combined with  MCD-based tracking algorithm, allowing us to characterize clusters with high precision.
Next, birth-death process of cluster is modeled and analyzed utilizing cluster lifetime and a first-order four-state Markov chain. 
Finally,  we develope a generalized CDL model  in accordance with 3GPP standard and validate its accuracy  by comparing  measurement and simulation  results, 
and the model has been proven to have improved performances than 3GPP.
This work enhances the understanding of 5G-R channels and will significantly contribute to the design, deployment, and optimization of 5G-R networks.

   \balance
   \bibliographystyle{IEEEtran}

   \nocite{*}
   
   \bibliography{IEEEabrv,ref}
\end{document}